# Vsep-New Heuristic and Exact Algorithms for Computing Graph Automorphism Group and Graph Isomorphism

Stoicho D. Stoichev

Department of Computer Systems, Technical University of Sofia
email: stoi@tu-sofia.bg

**Abstract.** Five new algorithms, named Vsep, are described. Four of them are for determining the generators, orbits and order of an undirected graph automorphism group. Vsep-e – exact, Vsep-orb and Vsep-sch – heuristic and Vsep-a automatically selects the optimal version among Vsep-e, Vsep-orb and Vsep-sch. The fifth algorithm, Vsep-is, is for finding an isomorphism between two graphs. Vsep-orb firstly finds heuristically the generators and orbits and then uses the exact one on the orbital partition for determining the order of the group. Vsep-sch differs from Vsep-orb in using the Schreier-Sims algorithm for determining the order of the group. A basic tool of these algorithms is the adjacency refinement procedure that gives finer output partition on a given input partition of graph vertices. The refinement procedure is a simple iterative algorithm based on the criterion of relative degree of a vertex toward a basic cell in the partition. A search tree is used in the algorithms - each node of the tree is a partition. All nonequivalent discreet partitions derivative of the selected vertices called a bouquet are stored in a coded form in a hash table in order to reduce the necessary storage – this is a main difference of Vsep-e with the known graph automorphism group algorithms. A new strategy is used in the exact algorithm: if during its execution some of the searched or intermediate variables obtain a wrong value then the algorithm continues from a new start point losing some of the results determined so far. The new start point is such that the correct results can be obtained. The proposed algorithms has been tested on the nauy&Traces benchmark graphs and compared with Traces, and the results show that for some graph families Vsep-e outperforms Traces and for some of the others Traces outperforms Vsep-e. The heuristic versions of Vsep are based on determining some number of discreet partitions derivative of each vertex in the selected cell of the initial partition and comparing them for an automorphism, i.e. their search trees are reduced. The heuristic algorithms are almost exact and are many times faster than the exact one. The heuristic algorithms are good choice for the user because of their smaller running times. Several cell selectors are used in Vsep, some of them are known and some are new. We also use a chooser of cell selector for choosing the optimal cell selector for the manipulated graph. The experiments show that the running time of Vsep algorithms does not depend on the vertex labeling.

## 1. Introduction

We assume some familiarity with the basics in the design and analysis of algorithms[1–3], combinatorial algorithms[4,5], graph theory and group theory [6–10]. We consider simple finite undirected graphs (without loops and multiple edges). The graph is denoted by G(V,E), where : V={1,2,3, . . . ,n} is the set of vertices and E - the set of edges (v, w), v, w ∈ V. The number $|E|$ of the edges of G we denote by $k$, $k=O(n^2)$. Our algorithms are applicable to any undirected graph including disconnected graphs but for them there is more efficient algorithm we do not describe here. The set of all vertices adjacent to a vertex x is denoted by *Adj(x)*. We use a static list representation of a graph (by two static arrays) because of its least required storage and fastest operation of finding all adjacent vertices of a given vertex compared with the adjacency matrix and the dynamic list representation.

*An isomorphism* [4–12] between two graphs $G_1(V_1, E_1)$ and $G_2(V_2, E_2)$ is called one-to-one correspondence (mapping) $y_i = f(x_i)$ between the vertices of the graphs ($x_i \in V_1$, $y_i \in V_2$, i=1,2,...,n, $n=|V_1|=|V_2|$) such that two adjacent (nonadjacent) vertices from one of the graphs correspond to two adjacent (nonadjacent) vertices from the another graph, i. e. every edge (non-edge) (p, q ) from graph $G_1$ corresponds to an edge (non-edge) (f(p),f(q)) from graph $G_2$ and vice versa. So, the isomorphism preserves the adjacency relation of vertices - this kind of bijection is commonly called "edge-preserving bijection".

Two graphs $G_1$ and $G_2$ are called *isomorphic* ($G_1 \cong G_2$) if there is at least one isomorphism between them. Otherwise they are *non-isomorphic,* G1!≈G2.

An *automorphism* [4–11,13] of a graph is an isomorphism of the graph onto itself. Or, an *automorphism* h of graph G(V,E) is called one-to-one correspondence $y_i=h(x_i)$ between the vertices of the graph ($x_i, y_i \in V$, i=1,2,...,n) that preserves the adjacency of the vertices, i. e. there is unique corresponding edge (non-edge) $(h(p),h(q)) \in E$ to each edge (non-edge) $(p,q) \in E$. *A fixed point x* of an automorphism *h* is called a vertex *x* for which $x = h(x)$. *Trivial automorphism* is an automorphism $h_0$ if each its vertex is a fixed point, $x_i=h_0(x_i)$, i=1,2,...,n and a *non–trivial automorphism* is an automorphism for which there is at least one pair of vertices *x,y* such that $y=h(x) \neq x$.

Two vertices $x_i$ and $y_i$ in the a graph G(V,E) are called *similar* (or symmetric) [6,7], $x_i \sim y_i$, when they are corresponding, $y_i=h(x_i)$, in some automorphism h. Otherwise, if $x_i$ and $y_i$ are not similar we use the notation $x_i !\sim y_i$. An automorphism *h* can be presented by two sequences $\Pi_a$ and $\Pi_b$ of graph vertex labels:

$a_1, a_2,..., a_i,... a_n=\Pi_a$

$b_1, b_2,..., b_i,... b_n=\Pi_b$, where the corresponding vertices are $a_i$ and $b_i=h(a_i)$, i=1,2,...,n.

The automorphism h may be shortly written in the form $\Pi_b=h(\Pi_a)$. The automorphism can be presented by *n!* pairs of rows - each row being derived from the other by transposing the positions of the pairs of corresponding vertices. The corresponding pairs of vertices can be set on to any place of the rows, but it is possible the place to depend on the sorting criterion which does not depend on the vertex labeling. Each automorphism can be written uniquely only with the permutation $\Pi_b$ if we assume that $\Pi_a = 1, 2, 3, ... n$. Even a simpler notation called *cycle notation* [10] is often used. In a cycle $(x_1, x_2, \ldots, x_i, x_{i+1}, \ldots, x_p)$ $x_i$ maps to $x_{i+1}$, $1 \leq i \leq p-1$ and $x_p$ maps to $x_1$. For example,

$$h=(1,8)(2,6,3,7)(4,5) = \begin{pmatrix} 1 & 2 & 3 & 4 & 5 & 6 & 7 & 8 \\ 8 & 6 & 7 & 5 & 4 & 3 & 2 & 1 \end{pmatrix}$$

Any automorphism can be written as the product of disjoint cycles and the product is unique up to the order of the cycles [10]. The cycles of length 1 are omitted.

*The operation function composition (or superposition)* [4–11,13] of two automorphisms $\alpha = \begin{pmatrix} ...X... \\ ...Y... \end{pmatrix}$ and $\beta = \begin{pmatrix} ...Y... \\ ...Z... \end{pmatrix}$ is a consecutive execution of α and β, i. e. $\gamma = \alpha.\beta = \begin{pmatrix} ...X... \\ ...Z... \end{pmatrix}$. This operation is usually called multiplication and is denoted by juxtaposition γ=α.β.

The set of all automorphisms of a given graph *G(V,E)* form *a graph automorphism group* (under the operation function composition of automorphisms) denoted *Aut(G),* shortly *A(G)* or *A* ) [4–11,13]. The trivial automorphism is the *identity* of the group – we denote it by *I*. The number of the automorphisms in *A(G)*, |*Aut(G)*|, is called *an order of the automorphism group*. *An order of an automorphism* is the order of the cyclic group generated by this automorphism. If the automorphism is written in cycle form then its order is the least common multiple of the cycle lengths[10].

The subset *gen(A)⊆A*, denoted *gen(A)* or *<g_1,g_2,...,g_d>*, is called *a generating set* [4,5,9,10,12] of the automorphism group if every automorphism $h \in A$ can be written as a finite product of elements $g_i, g_i^{-1} \in gen(A)$. The generating sets are not unique.

The subset of the graph vertices *y* similar to vertex *x* in any automorphism $h \in Aut(G)$, *y=h(x)*, is called an *orbit* of *x*, denoted *Orb(x,A)*: *Orb(x,A)={h(x)|h∈Aut(G)}* and *Orb(x,A)* is read as 'orbit of a vertex x under a group A'. By *Orb(A)* we denote the set of all orbits of the group *A*.

A graph with only one orbit of *Aut(G)* is called *transitive*, and a graph is called *rigid or asymmetric* if each of its vertices is in a separate orbit, i.e. *|Aut(G)|=1*. A stabilizer *A(x_1,x_2,...,x_i)* or *point-wise stabilizer* [5,10,13] is the subgroup of the automorphism group *A* that contains only the automorphisms with fixed points $x_1, x_2, \ldots, x_i$.

There is a theorem called *orbit-stabilizer theorem*[5–8] for computing the order |*A*| of the automorphism group of graph *G(V,E)*. We denote it O-S theorem. The theorem claims that *|A|=|A(x_1)|\*|Orb(x_1,A)|=|A(x_1)|\*d*, where *Orb(x_1,A)={ x_1, x_2,..., x_d}* is the orbit of vertex $x_1$ under the group *A*, and *A(x_1)* is a stabilizer of a representative $x_1$ of this orbit. In other words, *the order of the graph automorphism group is equal to the product of the order of a stabilizer of one of its vertices and the length of the orbit of this vertex*. Vsep algorithms and other known graph isomorphism and automorphism algorithms use the orbit-stabilizer theorem.

The stabilizer *A_i= A(x_1,...,x_i)* is called an *ancestor stabilizer* to the stabilizer *A_j= A(x_1,...,x_j)* if *j>i* and *A_j* is called a *successor (or descendant) stabilizer* to *A_i*. Obviously, the orbits of the successor stabilizer are subsets of the orbits of the ancestor stabilizer and the order of the successor stabilizer is a divisor of the order of the ancestor stabilizer.

Given a set *S⊆V* a *set-wise stabilizer* [10,12], denoted *Aut(G,S),* is a subgroup of *Aut(G)* defined by:*Aut(G,S)={α∈Aut(G) |x,α(x)∈S}*.

The graph isomorphism (*GI*) problem consists in deciding whether two given graphs are isomorphic, i.e., whether there is an isomorphism between them. The graph isomorphism problem belongs to the class of *NP* problems [4–9,11,12,14], and has been conjectured intractable, although probably not NP-complete.

It is neither known that this problem has polynomial time complexity nor to belong to the class of *NP–complete* problems. Its unknown complexity status is a theoretical and practical challenge. Two problems are called *polynomial-time equivalent* if there is a polynomial time algorithm that transforms one of the problems to another [4,7,9,10]. The problems [3,9,10] that are polynomial-time equivalent to graph isomorphism are called *graph isomorphism complete*. The following problems are graph isomorphism complete [15,16]: given two graphs G1 and G2: existence of isomorphism of the graphs; determine the isomorphism of the graphs if it exists; determine the numbers of the isomorphisms from G1 to G2; given a graph G: determine the generating set gen(Aut(G)), determine the orbits Orb(Aut(G)), determine the order |Orb(Aut(G))|. The problems of determining the orbits and order of the graph automorphism group *A* are also called *automorphism partitioning* and *automorphisms counting*, respectively. The generators, orbits and order of a graph G automorphism group *A* we denote by GOO(A) or GOO(Aut(G)) and by GO(A) - the generators and orbits of the group. A class of graphs is called graph isomorphism complete if the recognition of isomorphism for graphs from this class is a graph isomorphism complete problem. The following classes are graph isomorphism complete [16]: connected graphs, graphs of diameter 2 and radius 1, directed acyclic graphs, regular graphs, bipartite graphs without non-trivial strongly regular subgraphs, bipartite Eulerian graphs, bipartite regular graphs, line graphs, chordal graphs, regular self-complementary graphs, etc. However, there are special cases of the graph isomorphism problem with polynomial-time complexity: planar graphs (linear time), trees have a particularly simple algorithm, graphs of bounded degree, interval graphs, permutation graphs and convex graphs. At present it is not known a polynomial time algorithm for solving the graph isomorphism complete problems in the worst-case [14,17] – all known algorithms have *exponential* or *moderately exponential worst-case time complexity*. Graph isomorphism problems are of great practical and theoretical interest [3–5,12,14,17]. Recently, Laszlo Babai has claimed that the Graph Isomorphism problem can be solved in *quasipolynomial time* [18]. There are several practical algorithms (their names are shown below in parentheses) for graph isomorphism and graph automorphism group, due to Brendan McKay (nauty), Adolfo Piperno (Traces), William Kocay (Groups&Graphs), Schmidt and Druffel, Jeffrey Ullman; L.P. Cordella, P. Foggia C. Sansone and M.Vento (VF2), Tommi Junttila and Petteri Kaski (bliss), Hadi Katebi, Karem A. Sakallah, and Igor L. Markov (saucy), Jose Luis Lopez-Presa, Luis Nunez Chiroque, and Antonio Fernandez Anta (conauto), G. Tener and N. Deo (nishe), Nechepurenko [4], etc. There is a comparison in [13,19] on running times of the following tools: nauty, Taces, bliss, conauto and saucy. Their running time on random graphs is quite well but a major problem of these algorithms is their exponential time performance in the worst- case. There are two main generalizations of the graph isomorphism: subgraph problem (given two graphs determine if one of them is a subgraph to another) and largest common subgraph problem (given two graphs determine the common subgraph to both that has the maximum number of vertices or edges). These two problems are NP-complete and have many applications and the algorithms for them use the graph isomorphism and automorphism algorithms as basic tools. Another important problem related to graph isomorphism is [5,12]: Compute a complete invariant (certificate, signature) f for G, i.e. for all graphs G and H, G ≅ H ⇔ f(G) = f(H) (graph certificate identifies a graph uniquely up to isomorphism).
The graph isomorphism related problems (graph isomorphism itself, GOO(Aut(G))), subgraph isomorphism, largest common subgraph, graph certificate and canonization of a graph) arise in such fields as mathematics, chemistry, information retrieval, linguistics, logistics, switching theory, bioinformatics, and network theory [4,20].
Our goal is to develop exact and heuristic algorithms for determining GOO(Aut(G)), i.e. to solve the three problems by one algorithm with time complexity as lower as possible. In addition, our requirements to the heuristic algorithms are to give results equal to the results of the exact algorithm with the probability close to 1. There are a few heuristic algorithms for the graph isomorphism problem [21–24] – there is no access to their program codes. We propose three new heuristic algorithms (Vsep-orb, Vsep-hway, Vsep-sch) for GOO(Aut(G)) with much lower polynomial time complexity than the exact one. The experiments show that they are many times faster than the exact algorithm - even for difficult graphs with large sizes they give correct results. One of the first step in the exact algorithm is a call to a heuristic procedure for determining a representative of one of the smallest orbits of Aut(G) as a starting selected vertex – this way we speed up the exact algorithm and reduce the required storage.

## 2. Partitions and a refinement procedure

*An ordered partition* (or simply *partition*) Π or Π(G)
Π = $C_1C_2…C_i…C_p$ = $C_1 \cup C_2 \cup … \cup C_i \cup … \cup C_p$
of the vertices of graph G(V, E) [5,12,15] is a sequence of disjoint non-empty subsets of V whose union is V. Π =| $x_{1,1}, x_{1,2}, …, x_{1,k1}$ | … | $x_{i,1}, x_{i,2}, …, x_{i,ki}$ | … | $x_{p,1}, x_{p,2}, …, x_{p,kp}$| is a detailed presentation of Π, where $C_i$=| $x_{i,1}, x_{i,2}, …, x_{i,ki}$ |, $x_{i,j} \in V$, i=.1,…,p, j=1,…,ki. The subsets $C_1, C_2, … , C_p$, are called c*ells (classes, blocks)*.
We denote the number of the cells in a partition Π by |Π|. Two cells are called *adjacent* if there is at least one edge between their vertices, i.e. cells $C_i, C_j \in \Pi$ are adjacent if there is at least one edge (x,y), x∈ $C_i$, y∈ $C_j$. A cell with cardinality one is called *trivial (or singleton)*. The vertex of such a cell is said to be fixed by Π or it is called a *fixed point* of Π. A partition, of which each cell is trivial is called *discrete* or *numeration* having in mind that in fact it is a permutation that can be viewed as a graph vertices renumbering – vertex *i* corresponds to vertex *x* that is on the position *i* in the partition. By *NC(x, Π)* we denote the index of the cell *C* of Π that contains vertex *x*, *x ∈ C*. The position (index) of a vertex *x* in the partition (or in the cell) we denote by *pos(x)*. The *relative degree* ρ(x,$C_i$) of a vertex x∈$C_j$ toward a cell $C_i$ is equal to the number of vertices of cell $C_i$ adjacent to vertex x. We denote by v(x, Π) a *cell-degree vector* defined as *v(x, Π)=( ρ(x,$C_i$), i=1,…,p)*, p=|Π| – it is a vector whose components are the relative degrees of x to each cell in Π. We call a partition Π *stable (or equitable)* if the cell-degree vectors *v(x, Π)= v(y, Π)* for each two vertices *x, y ∈ $C_i$*, where $C_i$ is any cell in Π. We say that the partition $\Pi_2$ is *finer* than $\Pi_1$, written $\Pi_2 \leq \Pi_1$, if for every cell $C_i \in \Pi_2$ there exists a cell $C_j \in \Pi_1$ such that $C_i \subseteq C_j$. In order to get a finer partition $\Pi_2=D_1D_2...D_q$ when given $\Pi_1=C_1C_2...C_p$ , a refinement

procedure ($RP$, $\Pi_2 = RP(\Pi_1)$) is used, that assign to each vertex $x \in V$ a sorting criterion according by which the vertices of each class $C_i \in \Pi_1$ are sorted out in increasing or decreasing order of their criterions [4,5,12]. Often a sorting criterion is the relative degree $\rho(x,W)$ of any vertex $x \in \Pi$ toward some cell $W \in \Pi$. The *refinement procedure* that uses this criterion is called *adjacency refinement procedure*. We use only this version of the procedure. Example of another criterion for sorting is the number of the subgraphs of a given type (for example a triangle) that contain vertex *x*.

Two partitions $\Pi 1$ and $\Pi 2$ of the vertices of graph G(V,E) are called *compatible* [40] if: (1) $|\Pi 1| = |\Pi 2| = m$; (2) if $\Pi 1 = W_1 W_2 \ldots W_m$ and $\Pi 2 = U_1 U_2 \ldots U_m$, then for all $i \in [1:m]$, $|W_i| = |U_i|$; (3) for all $x, y \in V$, $NC(x,\Pi 1) = NC(y,\Pi 2)$ implies $v(x,\Pi 1) = v(y,\Pi 2)$. Similar definition is valid if $\Pi 1$ and $\Pi 2$ are for different graphs.

Two *partitions* $\Pi_1 = C_1 C_2 \ldots C_p$ and $\Pi_2 = D_1 D_2 \ldots D_q$ are called *equivalent* if there is an automorphism $\alpha \in Aut(G)$ such that $NC(x,\Pi_1) = NC(y,\Pi_2)$ for each pair of vertices $x$, $y = \alpha(x)$. In other words, the similar vertices are in cells with the same label of $\Pi_1$ and $\Pi_2$). Obviously for the equivalent partitions $\Pi_1$ and $\Pi_2$ we have: $p = q$ and $|C_i| = |D_i|$, $i \in [1:p]$. We denote the equivalent partitions by $\Pi_2 = \alpha(\Pi_1)$ or $\Pi_2 \approx \Pi_1$. Evidently the equivalence relation is transitive. Two discrete partitions $\Pi_1 = a_1 a_2 \ldots a_k \ldots a_n$, $\Pi_2 = b_1 b_2 \ldots b_k \ldots b_n$ of graph $G(V,E)$ vertices are called *equivalent* if they form an automorphism $h \in Aut(G)$, $b_k = h(a_k)$, $k = 1, 2, \ldots, n$. The testing if two discrete partitions $\Pi_1$ and $\Pi_2$ form an automorphism is a basic operation in our algorithms.

Given sequence of equivalent discrete partitions $\Pi_1, \ldots, \Pi_m$, we store one of them, for example $\Pi_1$ and the orbits derived by the automorphisms $\Pi_i = \alpha_i(\Pi_1)$, $i = 2, \ldots, m$.

*A partition-wise stabilizer A(G, Π) is defined by*: $A(G, \Pi) = \{\alpha \in Aut(G) \mid x, \alpha(x) \in C_i$, where $C_i$ is any cell of $\Pi\}$ [12]. This is a subgroup of Aut(G) such that each automorphism $\alpha \in Aut(G)$ belongs to Aut(G, Π) if to any vertex *x* of any cell of Π corresponds a vertex y=α(x) from the same cell. The orbits of *A(G, Π)* are subsets of the cells of *Π* as we'll see below. If $\Pi = \Pi_u$ then $A(G, \Pi_u) = Aut(G)$. Algorithms for determining the graph automorphisms and isomorphisms use very often the refinement procedure - each cell of its output partition contains at least one orbit of a graph automorphism group or its stabilizer. It is still not known a refinement algorithm that gives output partition each cell of which coincides with an orbit (*orbit partition or automorphism partition*) [20] on unit input partition.

One of the most efficient RP is the RP with a base cell. The RP with a base cell sorts (counting sort [1,2]) the vertices of any cell $C_j$ according to their relative degree *rdg(x,i)* toward a selected base cell $C_i$. The sorting continues until it reaches a partition $\Pi_2$ in which there is no cell that can be divided into subcells toward any base cell – such partition is called stable (equitable) [5,12] as we noted above: it holds the property $\Pi_1 = RP(\Pi_1)$. Vertices in every cell of the stable partition have the same sorting criterion - in our case, the same relative degree toward each cell. The base cell refinement algorithm has time complexity $O(\kappa . log\, n)$, where *k* and *n* are respectively the number of edges and the number of vertices of the graph [25,26]. The base cell RP refinement algorithm (Figure. 1) uses the counting sort that does not use a comparison operation – it sorts integers (the relative degree of the vertices) within the range 0 to some integer. In our algorithms for GOO(Aut(G)) we use only the adjacency refinement procedure with a base cell (Figure. 1) [25] and a W queue that contains the labels of all not selected cells as base cell. It differs from the known refinement algorithms [5,12,20] in the way the base cell is selected. After the current base cell ends sorting the adjacent cells as a new base cell is selected the first new smallest subcell and if there is no such subcell the label of the new base cell is taken from a W queue - the first cell label in W queue. There is a property that speeds up RP procedure: the label of the new largest subcell of C cell (adjacent to B cell) is not included in W queue if C cell label is not in W. There is a version of RP refinement algorithm with base cell that always takes the new base cell from the queue.

*The individualization-refinement* operation (denoted *IR*), used in the known GA algorithms, has two steps: individualization and refinement. Given an equitable partition $\pi$ and a vertex $x$ at the *individualization* step a new partition $\pi_1$ is obtained: the cell C($x$) of $\pi$ with index i is divided into 2 subcells: {$x$}-the first subcell, with index i and the second subcell {C($x$)\{$x$}} with index i+1, other cells of $\pi$ are not changed. At the *refinement* step the partition $\pi_1$ is refined with the refinement procedure RP obtaining a new equitable partition $\pi_2$=RP($\pi_1$) finer than $\pi_1$. Given a partition $\pi$ and a vertex $x$ we denote by IR($\pi$,$x$,br*cl*) the resulting partition $\pi_2$ from the application of IR operation on $\pi$ and $x$, where br*cl* is the number of cells of $\pi_2$.

| **Input:** graph G(V,E); $\Pi_1$ is the input partition on the graph vertices, W is a queue of some cells of $\Pi_1$, BRCL-the number of cells of $\Pi_1$ |
|---|
| **Output:** a better stable partition $\Pi_2$ ( $\Pi_2 \leq \Pi_1$), BRCL-the number of cells of $\Pi_2$ |
| S1: $\Pi_2 := \Pi_1$; |
| S2: Base cell B:= first cell in W; Delete the label B from W; |
| S3: **repeat**{each the loop execution is performed for different base cell B} |
| S4: Each cell C$\in \Pi_2$ adjacent to B is divided into subcells according to its relative vertex degrees toward B; |
| S5: **if** there are new subcells from S4 **then** B:= the label of the subcell with minimum length. **If** there are more cells with minimum length **then** the one with the smallest label is chosen. Put into W the labels of the new subcells in $\Pi_2$, excluding the label of one of the largest subcell if it is not in W; Delete the label B from W; |
| S6: **if** there are no new subcells from S4 **then** B:= first cell label in W; |
| S7: **until** there are no new subcells from S4 and W becomes empty |

**Figure. 1**. RP refinement procedure with base cell B

Cells in our algorithms are not consecutively labeled by 1, 2, … . The label NC(C$_j$,$\Pi$) of a cell C$_j$ in the partition $\Pi$ is determined by $NC(C_i, \Pi) = \sum_{j=1}^{i-1} |C_j| + 1$, i. e.

the label of the cell Ci (respectively of each of its vertices ) is the first vertex index in the cell, or it is greater by 1 than the number of the vertices in all cells preceding Ci in the partition. NC($x$,$\Pi$) denotes the label of the cell that contains vertex $x$. This way of labeling is time saving because changing the labels of a given cell does not cause change of the labels of other cells.

Two cells of a partition are called *non-trivially joined* (have non-trivial join, *non-uniformly joined*) if the number of edges between them is greater than 0 and less than maximum possible.

A *channel of a cell* C (new notion), denoted as *Ch(C))*, *is the number of the edges adjacent to any vertex in C cell*. A *channel of two cells* $C_1,C_2$, denoted as *Ch($C_1,C_2$))*, *is the number of all edges (x,y) between any vertex $x \in C_1$ and any vertex $\in C_2$*. Example: Let $\Pi=|1,8|4,5|2,3,6,7|=C_1\ C_2\ C_3$ be a partition on the vertices of the graph on Fig. 2. The channels of the cells are: *Ch($C_1,C_1$)=0, Ch($C_1,C_3$)=4, Ch($C_1$)=4, Ch($C_3,C_3$)=2, Ch($C_3,C_2$)=4, Ch($C_3$)=10, Ch($C_2,C_2$)=1, ChC($C_2$)=5*. Only the edges of cells with non-trivial join to C are included in *Ch(C)*. For example, the cells $D_5$ and $D_6$ of the partition $\Pi=|1|8|5|4|6, 7|2, 3|=D_1\ D_2\ D_3\ D_4\ D_5\ D_6$ of the vertices of the above graph have channels *Ch($D_5$)=1, Ch($D_6$)=1* since the edges incident to vertices from $D_5$ and $D_6$ and connected to vertices from the trivial cells $D_1\ D_2\ D_3\ D_4$ are not included – only the edge (6,7) is included in *Ch($D_5$)* and (2,3) in *Ch($D_6$)*.

*A channel vector of a cell $C \in \Pi$, ChVect(C), is a vector VC whose component VC(i) is equal to Ch($C_i$,C) channel where $C_i$ is a cell$\in \Pi$ with non-trivial join to C. A channel graph of a partition $\Pi$, ChG ($\Pi$), is an weighted graph with loops:* each vertex of ChG corresponds uniquely to a cell of $\Pi$ and its weight is the channel of the cell; each edge of ChG corresponds to a channel of the corresponding cells of $\Pi$ and the weight of this edge is equal to the weight of this channel. Similar notion is a *quotient graph* in [27]. *A selected non-trivial cell, SC($\Pi$),* of a partition $\Pi$ is the cell $C_j$, $|C_j|>1$, that is selected by a defined criterion (often it is called *target cell*, for example in [12]). The procedure that finds the target cell is called *cell selector*.

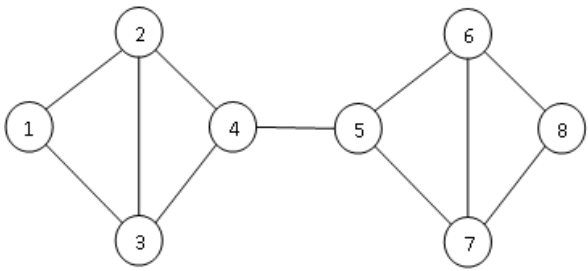

**Fig. 2. GIRA graph**

The selection of the target cell has strong influence on the search tree size (see the next section) and consequently, on the running time of the GA algorithms.

In Vsep algorithms, at given an equitable partition $\pi$ we use the following 5 cell selectors (their numbering is denoted by *izb*):

*izb*=1, MXNACS1: (a) it selects the nontrivial cell C of $\pi$ with maximal number of adjacent non-trivially joined cells to C, and (b) if there are several such cells it selects the first one of the smallest size;

*izb*=2, MXNACS2: it differs from (1) only in (b) selecting the first cell with largest size.

*izb*=3, MXVECTCHVAL: it selects the first nontrivial cell C of $\pi$ with the largest vector of relative degrees of C to other cells – the vector is considered as a number with leftist least significant digit.

*izb=4*, MXPRCHVAL: it selects the first nontrivial cell C of π with largest channel.
*izb=5*, ICLMXBRCL: it selects the first non-trivial cell C with maximal number of cells of the derived partition π(x), x ∈ π, obtained after IR.

To select the optimal cell selector we use a *cell selector chooser (CSLCh)* – it chooses the cell selector by three criteria: a) the smallest maximal level LMAX of the first path in the search tree., b) the smallest product of the sizes of all target cells of the first path in the search tree and c) the smallest product of the numbers of the cells of all partitions on the first path in the search tree.

The partitions in the developed algorithms can be:
1. *Unit partition* (denoted by $\Pi u = V = |1, \ldots, n|$) - all its vertices are in one cell $C_1$ and the cell number is $NC(i)=1$, $i=1,\ldots,n$.
2. *Equitable partition* - it is obtained as a result of the execution of the refinement procedure (RP) on given input partition (see the definition given above);
3. *Transformed partition* $\Pi^T$ is obtained from a stable partition Π by separating from SC(Π) given vertex x in a singleton cell. The difference between $\Pi^T$ and Π is that SC(Π) with cell label i, is divided into two cells in $\Pi^T$: cell C'={x} with label i containing vertex x and cell C''= SC(Π)\{x} with label i+1 containing the other vertices of SC(Π). The transformation operation is denoted by TR: $\Pi^T = TR(x, \Pi)$ and is called an *individualization* in many references.

The following theorem holds for the RP.

**Theorem 1**
*Given a graph G(V,E), two different equivalent partitions $\Pi_1$ and $\Pi_2$ on V and a non-trivial automorphism f∈Aut(G). Then, $NC(x,\Pi_a) = NC(y,\Pi_b)$ for each pair x, y=f(x), where $\Pi_a = RP(\Pi_1)$, $\Pi_b = RP(\Pi_2)$.*
*In other words, RP gives output partitions where the similar vertices remain in cells with the same label.*

This theorem is equivalent to the Theorem 7.1 in [5] – for equivalent partitions under an isomorphism of two graphs and the statement that the RP is invariant under an automorphism [28]. This is the reason for not giving here the proof of the theorem.

**Corollary 1 of Theorem 1** *Let Π be a partition on V of a graph G(V,E) with an automorphism f∈Aut(G) and let for each pair of similar vertices x, y=f(x) the property $NC(x,\Pi) = NC(y,\Pi)$ holds. Then, $NC(x,\Pi_c) = NC(y,\Pi_c)$, where $\Pi_c = RP(\Pi)$.*
Corollary 1 means that the similar vertices in any cell of a given input partition remain in one cell of the resulting partition of RP.

**Proof** Let $\Pi = \Pi_1 = \Pi_2$ in theorem 1. Then, we obtain $\Pi_c = \Pi_a = \Pi_b = RP(\Pi)$ and $NC(x,\Pi_c) = NC(y,\Pi_c)$ is obtained by replacing $\Pi_c$ in $NC(x,\Pi_a) = NC(y,\Pi_b)$. ♦

**Corollary 2 of Theorem 1** *The refinement procedure RP is invariant under the vertex orbits of a subgroup H⊆A=Aut(G(V,E)): if Orb(x,H)⊆C(x,Π) for each vertex x∈V of a given partition Π, then Orb(x,H)⊆C(x,$\Pi_c$) for a partition $\Pi_c = RP(\Pi)$. (In other words, if* each orbit of a given automorphism subgroup is a subset of a cell of the input partition $\Pi$, then the same is true for the output partition $\Pi_c=RP(\Pi)$. Obviously, this corollary holds for any stabilizer $A(x_1,x_2,...,x_i)$ of the automorphism group.)

**Proof** The condition $Orb(x,H) \subseteq C(x,\Pi)$ means that $NC(x,\Pi)=NC(y,\Pi)$, $y=f(x)$ for each $f \in H$, i. e. the condition of Corollary 1 of Theorem 1 holds for each $f \in H$ and from it follows $NC(x,\Pi_c)=NC(y,\Pi_c)$, $y=f(x)$ for each $f \in H$, i.e. $Orb(x,H) \subseteq C(x,\Pi_c)$. ♦

***Corollary 3 of Theorem 1*** $Orb(x,A) \subseteq C(x,\Pi_1)$ *holds for each vertex* $x \in V$ *of the initial partition* $\Pi_1=RP(\Pi_u)$ *of the graph* $G(V,E)$ *where* $\Pi_u$ *is the unit partition.*

**Proof** Let $H=A$ and $\Pi=\Pi_u$ for which obviously $Orb(x,A) \subseteq C(x, \Pi_u)$. Then, $Orb(x,A) \subseteq C(x,\Pi_1)$ follows from Corollary 2 where $\Pi_1=\Pi_c=RP(\Pi_u)$. ♦

### 3. Auxiliary algorithms

### 3.1. A1 algorithm for determining one discrete partition

The output of A1 algorithm (Figure 3) is a series of better partitions the last of which is discrete on given input partition $\Pi$. Using A1 we introduce new notions and prove a property used for speeding up our algorithms.
The following basic operations are performed in the A1 algorithm: $\{\Pi_1=RP(\Pi), L=1\}$; { determine $SC(\Pi_L)$ by some cell selector; select any vertex $X_L$ in the $SC(\Pi_L)$; $L:=L+1$; $\Pi_L:=IR(\Pi_{L-1}, X_{L-1})\}$; $L=2, \ldots$, LK, where $\Pi_{LK}$ is discrete and $LK<n$.
Given any input partition $\Pi$ the initial partition $\Pi_1=RP(\Pi)$ is determined first. In most cases we'll consider that the starting input partition $\Pi$ is the unit partition, $\Pi= \Pi u$.
A *selection level* is called the successive number $L$ of the refinement procedure *RP* calls for obtaining the partition $\Pi_L$ in A1 algorithm. $\Pi_L=\Pi(x_1,x_2, \ldots , x_{L-1})$ denotes the partition that is obtained by making L-1 selections $x_1,x_2, \ldots , x_{L-1}$ starting with the partition $\Pi_1$. The selected non-singleton cell in partition $\Pi_L$ is denoted by $SC(\Pi_L)$ or $SC(x_1,x_2, \ldots , x_{L-1})$. Five operations (rows 4, 5) are executed in the loop C: determining $SC(\Pi_L)$, a selection of vertex $x_L$ in $SC(\Pi_L)$, a move to next level (a forward move, L+1), obtaining the transformed partition $\Pi_L^T:=TR(x_{L-1},\Pi_{L-1})$, $\Pi_L:=RP(\Pi_L^T)$ (refinement with RP) – the last two operations are IR . Obviously, $\Pi_L$ is a stable partition. The determination of $SC(\Pi_L)$ is made by some cell selector. This execution continues until a discrete partition $\Pi_{LK}$ is obtained – its level of selection *LK* is called *terminal or final*. The process of obtaining the sequence $\Pi_1, \Pi_2, \ldots , \Pi_{LK}$ is called *series of forward moves (SFM)*.
The selection of a vertex also is made by some criterion but here we consider (for definiteness) the selection of the first vertex in SC( $\Pi_L$).
The five instructions (lines 4, 5) we denote by ***CSVSIR*** (cell selection, vertex selection, individualization and refinement) – cell selection in $\Pi_{L-1}$, vertex $X_{l-1}$ selection in $SC(\Pi_{L-1})$, L=L+1, individualization of $X_{L-1}$ ($\Pi_L:=IR(\Pi_{L-1},X_{L-1})$) and refinement of $\Pi_L^T$.
Each partition $\Pi_L$, L=1, 2, ... , LK is called *a partition-child* of $\Pi_{L-1}$ *(partition-parent)* and *a successor* of each of the partitions $\Pi_{L-1}, \Pi_{L-2}, \ldots , \Pi_1, \Pi_0$ *(partitions-ancestors of $\Pi_L$)*.

The selected cells SC (L), L=1, 2, . . ., L-1 for a given partition $\Pi_L$ are called *supporting SC* for $\Pi_L$, and the selected vertices $x_1, x_2, ... , x_{L-1}$ of $\Pi_L$ are called *base points or a base* of $\Pi_L$ *(or supporting points)*.

*A search tree (ST)* is an oriented tree whose root represents the initial partition $\Pi_1=RP(\Pi)$. Each node of the search tree corresponds to an equitable partition $\Pi_L$. Each arc coming out of a node $\Pi_L$ corresponds to a selected vertex from the SC($\Pi_L$). Thus, the partition $\Pi_{L+1}$ of a particular node $q$ on a level L+1 could be obtained from the partition $\Pi_L$ (node $p$ on a level L) by selecting a vertex x ∈ SC ($\Pi_L$), x being depicted on the edge p-q in ST. Let r = |SC($\Pi_L$)|, then each node $\Pi_L=\Pi(x_1,x_2, ... ,x_{L-1})$ with SC ($\Pi_L$) = $\{a_1 a_2 ... a_r\}$ has r nodes-children of level L+1:
$\Pi(x_1,x_2, ... ,x_{L-1},a_1)$, $\Pi(x_1,x_2, ... ,x_{L-1},a_2)$, ... , $\Pi(x_1,x_2, ... ,x_{L-1},a_r)$. Thus, each vertex in each SC($\Pi_L$), for each level L is selected in ST. Only in $\Pi_1$ one vertex is selected – the first vertex in SC($\Pi_1$).

The just described search tree ST may be called a *full search tree* in contrast to *the reduced*

| Input: a graph G(V,E), a partition $\Pi$ of V |
| --- |
| Output: Series of better partitions $\Pi_1 < \Pi_2 < ... < \Pi_L < .. < \Pi_{LK}$(discrete) |
| 1: L:=1; $\Pi_L$:=$\Pi$; $\Pi_L$:=RP($\Pi_L$); Output($\Pi_L$); |
| 2: **if** $\Pi_L$ is discrete **then** message; **return end if** |
| 3: **repeat** {loop C} |
| 4: determine SC($\Pi_L$) by some cell selector; select a vertex $X_L$ in the SC($\Pi_L$) |
| 5: L:=L+1; $\Pi_L$:=IR($\Pi_{L-1}$, $X_{L-1}$,brcl); Output($\Pi_L$); |
| 6: **until** brcl=n |

**Figure 3. A1 algorithm**

*search tree RST,* in which some vertices from SC($\Pi_L$) are not selected according to a certain criterion. In the reduced search tree, for each orbit of a stabilizer $A_{L-1}=A(x_1,x_2, ... ,x_{L-1})$ only one representative of the orbit($A_{L-1}$)∈SC($\Pi_L$) is selected – this is the first orbits vertex met during the traversal of the SC($\Pi_L$). This way of a selection of vertices is used in Vsep algorithms. It is based on the theorems proved below.

The reduced search tree used in Vsep algorithms is not explicitly presented in the algorithm – only the partitions of the path from the root to the current $\Pi_L$ are stored, i.e., the sequence $\Pi_1, ... , \Pi_L$.

3.2. **A2 algorithm for generating the full search tree**

**A2 a**lgorithm (Figure 4) generates all partitions of the search tree on a given input partition $\Pi$ of graph G(V,E) and the selection of a vertex $X_1$ ∈ SC(L) of the initial partition $\Pi_1$ = RP($\Pi$). The leaf nodes of ST are discrete partitions. A new partition $\Pi_L$ (a new node of ST) is obtained after each execution of the instruction 10. In A2, each of the vertices in every SC($\Pi_L$) is successively selected and after that the algorithm continues with a forward move until a discrete partition is obtained. There are two loops in A2: the loop C1 (lines 4-15) performs a *forward move* and the loop C2 (lines 9-13) performs the *backward move*. In the forward move, as in A1 algorithm, we begin from the partition $\Pi_L$ (on a level L) and selected vertex $x \in SC(\Pi_L)$, and we obtain successively the partitions $\Pi_{L+1}, \Pi_{L+2}, \ldots, \Pi_{LK}$ (discrete). The backward move (L:=L-1) is made when all vertices in the SC(L) are selected. We introduce a new notion – a *level of the last backward move*, denoted by LP and meaning the last value assigned to *L* by the instruction *10 (L:=L-1),* i. e. *LP* is the level to which the last backward move is made. We call LP *a working level* and *SC($\Pi_{LP}$)* - *a working cell* since the main operations of the algorithms are with the vertices in *SC($\Pi_{LP}$)*. Level LP is a significant variable for Vsep-e algorithm. A sequence of forward moves (SFM) starts from the level LP. A2 algorithm (as well as Vsep-e algorithm) generates the search tree in *preorder* [1,2]– first visiting the root (a vertex $\in SC(\Pi_L)$) then visiting the subtree of each vertex $\in SC(\Pi_{L+1})$ from left to right.

| |
|---|
| **Input**: a partition $\Pi$ of vertices of G(V,E) graph |
| **Output**: all partitions –nodes of the full search tree |
| 1: L:=1; $\Pi_L = \Pi$; $\Pi_L$:=RP($\Pi$); Output $\Pi_L$; |
| 2: **if** $\Pi_L$ is discrete **then** message ' the graph is rigid'; **return end if** |
| 3:   determine SC($\Pi_1$) by the cell selector; select a vertex $X_1 \in SC(\Pi_L)$; |
| 4:   **do** { C1: loop of the forward move} |
| 5:    L:=L+1; $\Pi_L$:=IR($\Pi_{L-1}, X_{L-1}$,brcl); Output $\Pi_L$; |
| 6:    **if** $\Pi_L$ is not discrete **then** |
| 7:      determine SC($\Pi_L$) by the cell selector; $X_L$:=first vertex in the SC($\Pi_L$) |
| 8:    **else** |
| 9:     **repeat**{ C2: loop of the backward move} |
| 10:       L:=L-1; |
| 11:       **if** L=1 **then return end if;** |
| 12:       $X_L$:=first unselected vertex in SC($\Pi_L$); |
| 13:      **until** $X_L \neq 0$; |
| 14:    **end if** |
| 15: **end do** |

**Figure 4. A2 algorithm**

*A bouquet* B($x_L$) or B($x_1, x_2, \ldots, x_L$) of a selected vertex $x_L \in SC(\Pi_L)$ is called the set of all mutually non-equivalent discrete partitions derived from the partition $\Pi_L$ and the selection of $x_L$, i.e. obtained from $\Pi(x_1, x_2, \ldots, x_L) = \Pi_{L+1}$. Only the first numeration obtained by Vsep-e algorithm is stored from the set of equivalent discrete partitions and is used as a representative in B($x_L$). The bouquet B($x_1, x_2, \ldots, x_L$) consists of the bouquets of the orbits
representatives of vertices in the SC($\Pi_{L+1}$), i. e. it consists of the bouquets B($x_1, x_2, \ldots, x_L, a_{L+1}^1$), B($x_1, x_2, \ldots, x_L, a_{L+1}^2$), . . . , B($x_1, x_2, \ldots, x_L, a_{L+1}^p$), where $a_{L+1}^1, a_{L+1}^2, \ldots, a_{L+1}^p$ are orbits representatives of the stabilizer A($x_1, x_2, \ldots, x_L$) in SC(L+1). Given the selected vertices $x_1, x_2, \ldots, x_{L-1}, x_L$ the following relation holds for the bouquets: $B(x_1) \subseteq B(x_2) \subseteq \ldots \subseteq B(x_{L-1}) \subseteq B(x_L)$, i.e., the bouquet of a selected vertex $x_L$ is included in the bouquets of all preceding selected vertices.

**Theorem 2** *Given a graph $G(V,E)$, a partition $\Pi$ on V, an automorphism $f \in Aut(G, \Pi)$, $NC(x,\Pi) = NC(y,\Pi)$ for each pair x, y=f(x) (a property- equal cell labels for each pair of similar vertices), two partitions $P_L=\Pi(a_1, a_2, \ldots, a_{L-1})$, $Q_L= \Pi(b_1, b_2, \ldots, b_{L-1})$ obtained by 2 series of executions of CSVSIR on the input partition $\Pi$ and $NC(a_i)=NC(b_i)$ for $b_i = f(a_i)$, $i=1,2, \ldots, L-1$ (equal cell labels of corresponding base vertices). Then, $NC(x, P_L) = NC(y, Q_L)$ for each pair x, y=f(x), $x \in P_L$, $y \in Q_L$ (the property holds for each pair of similar vertices).*

**Proof** (by the mathematical induction): (1). The **base case**: By assumption the property $NC(x, \Pi) = NC(y, \Pi)$ holds for the initial partition $\Pi$. In particular, $NC(a_1, \Pi)=NC(b_1, \Pi)$.

(2). The **inductive step**: Let the property holds for i-1: $NC(x, P_{i-1}) = NC(y, Q_{i-1})$ for each pair x, y=f(x). For $P_i^T$ and $Q_i^T$ the property $NC(x, P_i^T) = NC(y, Q_i^T)$ holds since by assumption $NC(a_{i-1})=NC(b_{i-1})$ for $a_{i-1}, b_{i-1}=f(a_{i-1})$ is true. For all other vertices of the cells $C(a_{i-1})$ and $C(b_{i-1})$ the cell labels are equal to $NC(a_{i-1})+1$ – this does not violate the property. For the vertices of other cells of $P_{i-1}$, $Q_{i-1}$ there is no change of cell labels in $P_i^T, Q_i^T$. Hence, according to Corollary 1 of Theorem 1 the property $NC(x, \Pi_i) = NC(y, P_i)$ also holds for $\Pi_i=RP(\Pi_i^T)$, $P_i=RP(P_i^T)$.

Since both the base and the inductive step have been proved, it has now been proved that $NC(x, \Pi_L) = NC(y, P_L)$ holds for *all* vertices x and y=f(x), $x \in P_L$, $y \in Q_L$. ♦

**Corollary of Theorem 2** *Given a partition $\Pi_L=\Pi(x_1, x_2, \ldots, x_{L-1})$. Then, $Orb(x, A_{L-1}) \subseteq C(x, \Pi_L)$ for each vertex $x \in V$ of the graph $G(V,E)$, $L=1, \ldots, LK$. (In other words, the orbits of $A_{L-1}$ are subsets of the cells of $\Pi_L$).*

**Proof** Let $b_i=a_i$, $i=1,2, \ldots, L-1$ in Theorem 2. Thus, these L-1 vertices become fixed points of automorphism f and $P_L=Q_L$. From $NC(x, \Pi_L) = NC(y, \Pi_L)$ for each pair x, y = f(x) and for each $f \in A_{L-1}=A(a_1, a_2, \ldots, a_{L-1})$ follows $Orb(x, A_{L-1}) \subseteq C(x, \Pi_L)$. ♦

Applying the corollary for L=1 we have $Orb(x, A_0)= Orb(x, A(\Pi)) \subseteq C(x, \Pi_1)$. Applying it to L=LK we obtain $Orb(x, A_{LK-1}) \subseteq C(x, \Pi_{LK})$ and $|Orb(x, A_{LK-1})| \subseteq |C(x, \Pi_{LK})|=1$ since $\Pi_{LK}$ is discrete, i.e., $|Orb(x, A_{LK-1})|=1$ and $A_{LK-1}=\{I\}$. Thus, for the series of partitions $\Pi_1, \ldots, \Pi_{LK}$ only the stabilizer $A_{LK-1}$ is known – the stabilizer of the first selected vertex $X_{LK-1}$ in $SC(\Pi_{LK-1})$.

**Theorem 3** *Given a graph $G(V,E)$, n=|V|, an automorphism $f \in Aut(G)$ with fixed points $a_1, a_2, \ldots, a_{j-1}$ and a partition $\Pi_j=\Pi(a_1, a_2, \ldots, a_{j-1})$, obtained after (j-1) successive applications of CSVSIR operation on $a_1, a_2, \ldots, a_{j-1}$ with starting input partition $\Pi$ and vertices*
*p, q $\in$ SC ($\Pi_j$), q=f(p).*
*Then, each numeration*

$\Pi'=\Pi(a_1, a_2, ..., a_{j-1}, p, a_{j+1}, a_{j+2}, ..., a_{LK-1})=|x_1|\ x_2|\ ...\ |\ x_i|\ ...\ |\ x_n|$,
derivative of $\Pi(a_1, a_2, ..., a_{j-1}, p)$ has a corresponding numeration
$\Pi''=\Pi(a_1, a_2, ..., a_{j-1}, q, b_{j+1}, b_{j+2}, ..., b_{LK-1})= |y_1|\ y_2|\ ...\ |\ y_i|\ ...\ |\ y_n|$,
derived from $\Pi(a_1, a_2, ..., a_{j-1}, q)$, such that $y_i=f(x_i)$, $i=1, 2, ..., n$ and
$NC(x_i, \Pi')=NC(y_i, \Pi'')$.
(Note: There is an equivalent theorem of B. McKay - theorem 2.15 in [12]).
.
**Proof** Let we consider two executions (labeled I and II) of the operation CSVSIR on starting input partition $\Pi$ of graph $G(V, E)$. The first $j-1$ selections $a_1, a_2, ..., a_{j-1}$ are equal for both executions and the resulting partitions are equal to $\Pi_j$. Under the conditions of the theorem, there are vertices $p, q=f(p)$ in the cell SC ($\Pi_j$).
Let the $j^{th}$ selection be $p$ in the first execution, and the $j^{th}$ selection be $q$ in the second execution, i.e. the obtained partitions are: for execution I – $\Pi_I=\Pi(a_1, a_2, ..., a_{j-1}, p)$, for execution II – $\Pi_{II}=\Pi(a_1, a_2, ..., a_{j-1}, q)$, for which the conditions of Theorem 2 hold: $a_i=f(a_i)$ for $i = 1, 2, ..., j-1$ and $q=f(p)$. Therefore, according to Theorem 2, the similar vertices are in cells with the same label, and thus the vertex $a_{j+1}$ in SC ($\Pi_I$) will correspond to the vertex $b_{j+1}= f(a_{j+1})$ in SC ($\Pi_{II}$). So, the selection $a_{j+1}$ is possible in execution I and the selection $b_{j+1} = f(a_{j+1})$ is also possible in execution II. The conditions of Theorem 2 also hold for these selections and at the $(j+2)^{th}$ selection similar vertices can be selected again, i. e. $a_{j+2}$, in the execution I and $b_{j+2} = f(a_{j+2})$ in the execution II. This process continues until the last selections $a_{LK-1}$, $b_{LK-1}=f(a_{LK-1})$ in both executions have been done – after these selections the partitions will be discrete. This means that each numeration
$\Pi(a_1, a_2, ..., a_{j-1}, p, a_{j+1}, a_{j+2}, ..., a_{LK-1})= |x_1|\ x_2|\ ...\ |\ x_i|\ ...\ |\ x_n|=\Pi'$
has a corresponding numeration
$\Pi(a_1, a_2, ..., a_{j-1}, q, b_{j+1}, b_{j+2}, ..., b_{LK-1})= |y_1|\ y_2|\ ...\ |\ y_i|\ ...\ |\ y_n|=\Pi''$, such that
$NC(x_i, \Pi')=NC(y_i, \Pi'')$, $y_i=f(x_i)$, $i=1, 2, ..., n$. ♦
There are four obvious corollaries of Theorem 3:

***Corollary 1 of Theorem 3*** *Under the conditions of Theorem 3 the discrete partitions of type $\Pi''$, successors of the partition $\Pi(a_1, a_2, ..., a_{j-1}, q)$, do not find new automorphisms (new similar vertices). Consequently, it is not necessary to determine them if we pre.liminarily know the discrete partitions of type $\Pi'$ - successors of the partition*
$\Pi(a_1, a_2, ..., a_{j-1}, p)$.

***Corollary 2 of Theorem 3*** *The bouquets of two similar vertices in a given $SC(\Pi_L)$ are of the same size.*
    This statement is obvious because to each numeration of the one bouquet uniquely corresponds a numeration of the other bouquet.

***Corollary 3 of Theorem 3*** *To determine whether two vertices X and Y in $SC(\Pi_L)$ are similar we need to know the bouquet of one of the vertices, say **B(L, X)**, and generate one numeration $n_1$ derived of a selection Y in SC(L) and compare $n_1$ with the numerations $\in$ **B(L, X)**. Even more, **B(L, X)** should not contain equivalent numerations* *because of the transitivity of the equivalence: if $n_1$ is equivalent to one of them it is equivalent to the another.*

***Corollary 4 of Theorem 3*** *The bouquet $B(L, X)$ contains all bouquets derived from each representative of an orbit in $SC(\Pi_{L+1})$.*

.Important conclusions follow from Theorem 3 and its corollaries. There are three possibilities to determine whether two vertices *X* and *Y* in $SC(\Pi_L)$ are similar under $A(x_1, \ldots, x_{L-1})$: (a) The bouquet *B(L, X)* of the vertex *X* should be stored and for the vertex *Y* we should generate only one numeration and compare it with the numerations $\in B(L, X)$ - this version is used in Vsep-e algorithm; (b) One numeration should be stored for vertex *X* and the whole bouquet for vertex *Y* should be generated. This version is used in Nauty [12] and in the most of the known algorithms; (c) Two bouquets *B(L, X)* and *B(L, Y)* are partially generated and their numerations are compared for determining an automorphism (with a certain probability) that maps *X* to *Y*. This probability might be near to 1 if we choose an appropriate selection of the bouquets size. This is the basis for the heuristic algorithms described in Section 5.

| version | *NS*-number of stored numerations | *NG*-number of generated numerations | *NC* - number of comparisons of numerations |
|---|---|---|---|
| a | *m* | *m+q-1* | *c.m(q-1)* |
| b | *1* | *m+m(q-1)=q.m* | *c.m(q-1)* |

**Table 1**

Let's compare versions (a) and (b) (Table 1). Let *m=|B(L, X)|* and let's consider that the numerations of the bouquets are stored in a hash table with a maximum number *c* of collisions of some hash function (characteristic of the numeration) we'll explain below. Let's also consider the worst-case – a rigid regular graph for *L=1* and $|SC(\Pi_1)|=q$ – in this case all vertices in $SC(\Pi_1)$ are not similar each other. This is the worst-case since: (i) for *L=1* the bouquets have the larger size than the bouquets for *L>1* and (ii) the bouquets for rigid graphs are full – each vertex at each level is selected. The advantage of a version (b) is a low storage – only one numeration is stored and the disadvantage of version (a) is the large required storage – the whole bouquet of size *m* for the first vertex $x_1 \in SC(\Pi_1)$ is stored. A version (a) is faster since the number of the generated numerations is smaller: *NG=m+q-1*. In this case the bouquet of the first vertex $x_1 \in SC(\Pi_L)$ is generated and stored and for each of the other *q-1* vertices only one numeration is generated - totally *NG =m+q-1*. In case of version (b) for each vertex $x \in SC(\Pi_1)$ all *m* numerations of bouquet *B(1, x)* are generated, i.e. *NG= m.q* – we suppose that the size of each bouquet is *m* or *m* is the largest size. Since *m.q >> m+q-1* version (a) is many times faster than version (b). The number (*NC*) of the comparisons of numerations is *c.m(q-1)* for both versions. In version (a) one numeration for each vertex in SC(1) is compared with *c.m* numerations of B($X_1$). In

version (b) the only stored numeration $n_1$ derived from the selection $X_1$ is compared with *c.m* numerations of each bouquet *B(X), x ∈ SC(1), x≠x_1*.

Examples:
- Graph A29_1 (rigid regular graph from [29]): n=29, m=14 (this size is for each vertex in $SC(\Pi_1)$), q=n=29; NG(a)=m+q-1=42, NG(b)=q*m=29*14=406 (in the brackets is the numbering of the version). We see the big difference between the .numbers of generated numerations of the two versions.
- Graph G1275 (Rigid affine plane of order 25, received from R. Mathon in private communication): n=1275, bipartite graph with k=625*26=650*25=16250; m=8322060; q=625; NG(a)=m+q-1=8322060+625-1=8322684; NG( b)=q*m = 625*8322060 =5201287500. In this case the difference between NG (a) and NG (b) is impressive.

## 4. S-code of a partition and storing the bouquets

We propose new code, named S-code, of a partition of the graph vertices. The partition code is a number depending on the labels, sizes of the partition cells and the number of the edges between the cells. S-code is used for reducing the time of comparing the partitions in the graph isomorphism and automorphism algorithms. The code of a given partition can be computed directly from the partition and the graph representation or from the code of the parent partition and the differences between the partition and its parent partition.

In our algorithms a large number of discrete partitions (numerations) of graph vertices are generated and stored. The length of each partition is *n* (*n* is the number of the graph vertices). One way of reducing storage requirements is the coding of partitions. To every partition is assigned a code (a number, characteristic value). The codes of two partitions are compared (instead of comparing the corresponding partitions) and if they are equal then the partitions are compared to determine if they form an automorphism. In this case the partitions have to be regenerated using the stored base *B, p=|B|* of the partition and applying the IR operation *p* times. The cardinality *p* of the base is many times less than *n*. Let's consider the storing of a partition $\pi_L$ obtained from the start partition $\pi_0$ by applying the IR operation successively. There are 3 ways of storing the partition $\pi_L$: **a)** storing the partition itself, i.e. *n* numbers are stored; b) storing the base $B(\pi_L)$ and the code $c(\pi_L)$, i.e. *p+1* numbers are stored – one for the code and *p* for the base. This way the amount of the stored information is reduced from n to p+1 numbers, where p << n. In this case a regeneration of $\pi_L$ is made when using of $\pi_L$ is needed; c) storing the code $c(\pi_L)$ and a polynomial code of *B*, i.e. only 2 numbers are stored but a regeneration of both $\pi_L$ and *B* is needed. In our implementations the version (b) is used.

The requirements for the code are:
i) The codes of the equivalent partitions have to be equal;
ii) The splitting ability of the code has to be maximal. This means that the number of not equivalent partitions with equal codes have to be minimal (minimum collisions);
iii) The computation of the code should have minimal number of operations (easy to compute);
We have examined a few versions of coding and the code with the best satisfaction of the requirements is the following:

$$code(\pi) = \sum_{(x,y) \in E} L(C(x)).L(C(y)), \text{ where:} \tag{4.1}$$

π - the adjacency refinement partition of the vertices of graph G(V,E), (x,y)- an edge of the graph, C(x),C(y) – the cells of the vertices x,y∈V and L(C(x)), L(C(y)) – the labels of the cells C(x),C(y). The label of a cell is the index of the first vertex in the cell representing a partition as an array.

The following theorem proves that the requirement (i) holds for the code (4.1):

**Theorem** Given a graph G and $\pi_1 \equiv \pi_2$, then $code(\pi_1) = code(\pi_2)$.

Proof Each edge (x,y)∈E(G) has unique image (α(x),α(y))∈E(G) for an automorphism α defining the equivalency of $\pi_1, \pi_2$. Moreover, C(x) = C(α(x)), C(y) = C(α(y)) – the similar vertices are in namesake cells. Therefore,
L(C(x)) = L(C(α(x))), L(C(y)) = L(C(α(y))) and
L(C(x)).L(C(y)) {∈ $code(\pi_1)$ }= L(C(α(x))).L(C(α(y))) {∈ $code(\pi_2)$}.
Consequently, $code(\pi_1) = code(\pi_2)$, since the last equation holds for each edge ∈E(G).

Evidently, the time complexity of computing the code by (4.4.1) is $T=k=O(n^2)$ multiplications (k-the number of the graph edges) since $k= O(n^2)$. The code of π can be computed directly by (4.4.1) or indirectly by the code of the parent partition of π. The maximal value of the code Max(Code(π)) is obtained for a discrete partition π of a complete graph on n vertices (since its number of vertices is largest) :

$$\text{Max}(code(\pi)) = \sum_{i=1}^{n-1} i(i+1+i+2+\ldots+n-1+n) = \sum_{i=1}^{n-1} i \frac{(n+i+1)(n-i)}{2} =$$

$$= \frac{1}{2}((n^2+n)\sum_{i=1}^{n-1} i - \sum_{i=1}^{n-1} i^3 - \sum_{i=1}^{n-1} i^2) =$$

$$= \frac{1}{2}((n^2+n)\frac{n(n-1)}{2} - \frac{n^2(n-1)^2}{4} - \frac{n(n-1)(2n-1)}{6}) = \frac{n(n-1)(n+1)(3n+2)}{24} \tag{4.2}$$

The second multiplicand of the first expression in (4.4.2) is a sum of an arithmetic progression.

*Example***:** Let's consider the graph in Figure 5 and the series of partitions and their codes:

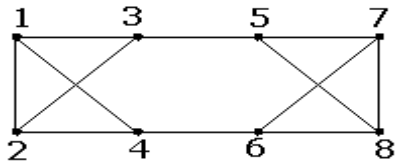

**Figure 5. G8 graph**

$\pi_0 = \pi(0) = |1,2,3,4,5,6,7,8| = C_1^0$; Code($\pi_0$)=12; $\pi_1 = \pi(1)$ = RP (|1| 2,3,4,5,6,7,8|) =

$|1|7,8|5,6|3,4|2| = C_1^1 C_2^1 C_3^1 C_4^1 C_5^1$ ; Code($\pi_1$)=208; $\pi_2 = \pi(1,7)$ = RP (|1|7|8|5,6|3,4|2|) =

$|1|7|8|5,6|3,4|2|) = C_1^2 C_2^2 C_3^2 C_4^2 C_5^2 C_6^2 C_7^2$ Code($\pi_2$)=218; $\pi_3 = \pi(1,7,5)$ =

RP(|1|7|8|5|6|3,4|2|) =|1|7|8|5|6|4|3|2|) = $C_1^3 C_2^3 C_3^3 C_4^3 C_5^3 C_6^3 C_7^3 C_8^3$ ; Code($\pi_3$)=234. We illustrate only how Code($\pi_1$) is computed (Table 2).

| edge (x,y) | 1,2 | 1,3 | 1,4 | 2,3 | 2,4 | 3,5 | 4,6 | 5,7 | 5,8 | 6,7 | 6,8 | 7,8 | Code ($\pi_1$) |
|---|---|---|---|---|---|---|---|---|---|---|---|---|---|
| L(C(x))* L(C(y)) | 1*8 =8 | 1*6 =6 | 1*6 =6 | 8*6 =48 | 8*6 =48 | 6*4 =24 | 6*4 =24 | 4*2 =8 | 4*2 =8 | 6*2 =12 | 6*2 =12 | 2*2 =4 | 208 |

**Table 2**

## 5. The exact Vsep-e algorithm

In this and the following section we describe the proposed algorithms some of which were preliminary presented in arXiv [30].

### 5.1. Basics of the algorithm

We need the following theorem for the reasoning of Vsep-e algorithm.

***Theorem 4*** Given: $A=Aut(G(V,E))$, an orbit $Q=Orb(x_1,A)$ of vetex $x_1 \in V$ and a generating set $gen(A(x_1))$ of a stabilizer $A(x_1)$. Then, there is a tower of acsending subgroups of $A$ $A(x_1)=A^{(1)} \subset A^{(2)} \subset \ldots \subset A^{(i)} \subset \ldots \subset A^{(m)}=A$, such that: (a) $A^{(i)}$ is a proper subgroup of $A^{(i+1)}$, $1 \leq i \leq m-1$; (b) $gen(A^{(i)})=\{gen(A^{(i-1)}), \alpha_i\}$, $2 \leq i \leq m$, where the autmorphism $\alpha_i$ is such that $x_i=\alpha_i(x_1)$; (c) $|A^{(i)}| \geq 2.|A^{(i-1)}|$, $|Orb(x_1, A^{(i)})| \geq 2.|Orb(x_1,A^{(i-1)})|$, $2 \leq i \leq m$; (d) $m$ is the minimal value of $i$ such that $Orb(x_1, A^{(m)})=Q$.
{Note: Evidently, the consequence of (b) is $gen(A)=gen(A^{(m)}))=\{gen(A(x_1)), \alpha_2, \alpha_3,\ldots,\alpha_m\}$. Theorem 4 can be considered as an extension or modification of Theorem 5 of C. M. Hoffmann in [10], page 25}

***Proof*** (inductive, it follows the proof of Theorem 5 in [10]). Let's construct a tower of ascending subgroups of $A$: $A(x_1)=A^{(1)} \subset A^{(2)} \subset \ldots \subset A^{(i)} \subset \ldots \subset A^{(m)}=A$, such that $A^{(i)}$ is a proper subgroup of $A^{(i+1)}$, $i=1,2,\ldots,m-1$; $m$ is finite since $A$ is a finite group. For $i=1$ we have $A^{(1)}=A(x_1)$ and $gen(A^{(1)})=gen(A(x_1))$, known. Assume inductively that $A^{(i)}$ is a proper subgroup of $A$ and let we have orbits $Q'= Orb(x_1, A^{(i-1)}) \neq Q''= Orb(x_1, A^{(i-1)})$, $Q', Q'' \subset Q$. Then, we'll have an automorphism $\alpha_i \in A \setminus A^{(i-1)}$ such that $x_i=\alpha_i(x_1)$ and $Orb(x_1, gen(A^{(i)})) \supseteq Q' \cup Q''$. Thus, $\alpha_i$ is a generator of $A$ and $A^{(i)}$ since it does not belong to $A^{(i-1)}$, i.e. $gen(A^{(i)})=gen(A^{(i-1)}) \cup \alpha_i$ and therefore $|A^{(i)}| \geq 2.|A^{(i-1)}|$ (since at least one new coset appear in the partitioning of $A^{(i)}$ into cosets of $A^{(i-1)}$) and $|Orb(x_1,A^{(i)})| \geq 2.|Orb(x_1,A^{(i-1)})|$ (from the orbit-stabilizer theorem: $|Orb(x_1,A^{(i)})|=|A^{(i)}|/.|A(x_1)| \geq |Orb(x_1,A^{(i-1)})|=||A^{(i-1)}|/A(x_1)|$).
The first value of $i$ when $Orb(x_1,gen(A^{(i)}))=Q$ and $A-A^{(i)}=\varnothing$ is $i=m$ and consequently $gen(A)=gen(A^{(m)}))=\{gen(A(x_1)), \alpha_2, \alpha_3, \ldots, \alpha_m\}$. ♦

Theorem 4 gives us the idea how to find the generators of a group if we know an orbit Q of the group and the generators of a stabilizer of a representative $x_1$ of this orbit. This is done by traversing the orbit step by step. At each step we find one new generator of a new subgroup of $A$ knowing the generators and the orbits of the previous subgrpoup of $A$. Before the first step the previous subgroup is equal to $A(x_1)$ with its orbits and generators.

Visiting each vertex *x* of the orbit, $x \neq x_1$, we select *x* only if *x* is not similar to $x_1$ under the previous subgroup. Thus a new automorphism-generator that maps *x* to $x_1$ and new orbits for a new subgroup are found. The new subgroup is a proper supergroup of the previous subgroup.. At each step the generators, orbits and order of the current group A (subgroup of A) are defined by the position of the selected vertex *x*. This process stops when the orbit of $x_1$ under the new subgroup becomes equal to the given orbit *Q*. The described process is presented in Table 3. The selected vertex $x_i$, i=2, . . . ,m is the first vertex after $x_{i-1}$ in Q that is not similar to $x_1$ under $A^{(i-1)}$. The generators of $A(x_1)$ are called *proper generators* of *A* and the generators $\alpha_2, \alpha_3, \ldots, \alpha_m$ - *mutual generators* of *A*. The following corollary is obvious:

| Selected vertex x | Visited vertices | Mutual generators | gen(A) | $|Orb(x_1,A)|$ | $|A|$ | A |
|---|---|---|---|---|---|---|
| $x_1$ | $\{x_1\}$ | - | gen($A(x_1)$) (proper generators of A) | $1=|\{x_1\}|$ | $|A(x_1)|$ | $A^{(1)}=A(x_1)|$ |
| $x_2$ | $\{x_1:x_2\}$ | $x_2=\alpha_2(x_1)$ | gen($A(x_1)$)$\cup \alpha_2$ | $\geq 2$ | $\geq 2.|A(x_1)|$ | $A^{(2)}$ |
| $x_3$ | $\{x_1:x_3\}$ | $x_3=\alpha_3(x_1)$ | gen($A(x_1)$)$\cup \alpha_2 \cup \alpha_3$ | $\geq 4$ | $\geq 4.|A(x_1)|$ | $A^{(3)}$ |
| $x_4$ | $\{x_1:x_4\}$ | $x_4=\alpha_4(x_1)$ | gen($A(x_1)$)$\cup \alpha_2 \cup \alpha_3 \cup \alpha_4$ | $\geq 8$ | $\geq 8.|A(x_1)|$ | $A^{(4)}$ |
| . . . | . . . | . . . | . . . | . . . | . . . | . . . |
| $x_i$ | $\{x_1:x_i\}$ | $x_i=\alpha_i(x_1)$ | gen($A(x_1)$)$\cup \alpha_2 \cup \ldots \cup \alpha_i$ | $\geq 2^{i-1}$ | $\geq 2^{i-1}.|A(x_1)|$ | $A^{(i)}$ |
| . . . | . . . | . . . | . . . | . . . | . . . | . . . |
| $x_m$ | $\{x_1:x_m\}$ =Q | $x_i=\alpha_m(x_1)$ | gen($A(x_1)$)$\cup \alpha_2 \cup \ldots \cup \alpha_m$ | $|Q|\geq 2^{m-1}$ | $\geq 2^{m-1}.|A(x_1)|$ | $A^{(m)}$=A |

**Table 3**

*Corollary of Theorem 4* The number of the mutual generators of the graph automorphism group A toward the stabilizer $A(x_1)$ is $m=1+log_2|Orb(x_1, A)|$.

Based on Theorems 3 4 and O-S theorem we build a new algorithm called A3. Actually, Theorem 4 tell us that there is a generator $x_i=\alpha_i(x_1)$ if $x \neq x_1$ is not similar to $x_1$ under the current group but according to Theorem 3 to determine this generator we should know the bouquet $B(x_1)$ and one numeration derivative of x and to compare them. Knowing the partition $\Pi_1$ and $SC(\Pi_1)$ according to Theorem 4 to traverse the unknown Q orbit we should traverse its superset, $SC(\Pi_1) \supseteq Q$. Thus, we come to the idea of A3 algorithm (Figure 6). We denote by *FRPO(X)* – all first representatives of the orbits of A positioned in $SC(\Pi_1)$ before the selected vertex *X* and by *BFRPO(X)* – the bouquets of *FRPO(X)*. Then, the bouquet B may be considered as a union of the bouquets *BFRPO(X)*.

A3 algorithm determines GOO(A) of the graph automorphism group $A=Aut(G,\Pi_1)$ ) and the bouquet of each representative of an orbit $Orb(A) \in SC(\Pi_1)$ given $GOO(A(X_1))$ of a stabilizer $A(X_1)$ of the first vertex $X_1 \in SC(\Pi_1)$ and the bouquet $B(X_1)$.

According to Theorem 3 and 4 we'll determine GOO(A) and the bouquet of each representative of an orbit $Orb(A) \in SC(\Pi_1)$ visiting each vertex $x \in SC(\Pi_1)$ positioned after $X_1$. Before the traversal of $SC(\Pi_1)$ we have $GOO(A)=GOO(A(X_1))$. Each visited vertex $x \in SC(\Pi_1)$ is selected if it is not similar to any previous vertex $\in SC(\Pi_1)$ under the current group $A$. Then, we determine if there is an automorphism $\alpha$, $x=\alpha(y)$, where $y$ is one of the roots of numerations in $B$. This is made (according to Theorem3) by comparing the first numeration $\Pi_{LK}$ derived from the selection x with the numerations $\in B$. Two cases are possible: (a) If there is an automorphism $\alpha$ then it belongs to a new subgroup of A since it unites the orbits of $x$ and $X_1$, i.e., $gen(A): = gen(A) \cup \{\alpha\}$; (b) If $\Pi_{LK}$ is not equivalent to any numeration $\in B$ then $x$ becomes a representative of a new orbit and a new search tree ST(x) is built - it determines gen(A(x)) and B(x). If during the generation of ST(x) an automorphism is discovered and if it unites the orbits of A then, it is also a generator for A since A(x)$\subseteq$A. Thus, after case (b) for the next selected vertex we should compare the numeration $\Pi_{LK}$ derived from the selection of $x$ with the numerations of all known bouquets B.

| | |
|---|---|
| **Input:** a graph G; a partition $\Pi_1=RP(\Pi)$ for a given input partition $\Pi$; $SC(\Pi_1)$; a vertex $XF \in SC(\Pi_1)$ and its index in $SC(\Pi_1)$; $GOO(XF)$: the generating set $gen(A(XF))$ of a stabilizer $A(XF)$, the orbits of $A(XF)$ and $|A(XF)|$ and the bouquet $B(XF)$; the bouquets $BFRPO(XF)$, i.e., the bouquet of each first representative $y$ of an orbit of $A$ in $SC(\Pi_1)$ with index $(y) < index (XF)$; each representative $y$ and $XF$ has no similar vertices in a position in $SC(\Pi_1)$ before it. | |
| **Output:** The generators, the orbits and the order (shortly GOO) of the graph automorphism group $A=Aut(G, \Pi_1)$; the bouquet B of each representative of an orbit of $A$ in $SC(\Pi_1)$. | |
| 1. | orbits of A:=orbits of A(XF); gen(A):=gen(A(XF)); |(A)|:= |A(XF)| |
| 2. | X:=XF; |
| 3. | **Do** |
| **4.** | select next $X \in SC(\Pi_1)$; **if** there is no selected vertex X **then exit;** |
| 5. | determine a numeration *n1* by SFM(X); |
| 6. | compare the numeration n1 with the numerations of the BFRPO(X); |
| 7. | **if** n1 is equivalent to some numeration from BFRPO(X), i.e. there is new automorphism $\alpha$ mapping X to some vertex from FRPO(X) **then** |
| **8.** | gen(A):= gen(A)$\cup \alpha$; recompute the orbits of A; |A|:=|Orb(XF)|.|A(XF)| |
| 9. | **else** {n1 is not equivalent to any numeration from BFRPO(X), i.e. there is no new automorphism mapping X to some vertex from FRPO(X)} |
| 10. | build the search tree ST(X) for determining GOO(A(X)) and the bouquet B(X) |
| 11. | **end;** {if} |
| 12. | **end do** |

**Figure 6. A3 algorithm**

To determine the orbit of the vertex $X_1$ it is sufficient to do the above comparison only with the bouquet $B(X_1)$ but we do this comparison with the numerations of all bouquets of B since we store them and the earlier finding of a generator is better since we can use it earlier. If the found automorphism $\alpha$, $x=\alpha(y)$, $y \neq X_1$ unites orbits of the current $A$ it is a generator of $A$.

After each of both cases has been handled the traversal of $SC(\Pi_l)$ continue by selection of a new vertex x. Thus, after the traversal of $SC(\Pi_l)$ all vertices$\in Orb(X_1, A)$ will be visited and the bouquet of each first representative $x_i \in SC(\Pi_l)$ of an orbit of $A$ and $GOO(A)$ will be determined (according to Theorem 4).

During the execution of A3 algorithm (as in Theorem 4) a tower of ascending subgroups of $A$ are built implicitly: $A(x1)=A^{(0)} \subset A^{(1)} \subset \ldots \subset A^{(i)} \subset \ldots \subset A^{(m)}=A$, such that $A^{(i)}$ is a proper subgroup of $A^{(i+1)}$, $0 \leq i \leq m-1$.

A3 algorithm is applied differently for $\Pi_L$ partition of level $L=1$ and $L>1$. If L=1 then $GOO(X_1)$ and $B(X_1)$ are determined only for the first vertex $X_1 \in SC(\Pi_l)$ and for other orbit representatives in $SC(\Pi_l)$ only one numeration is generated and compared with $B(X_1)$. This means that instruction in line 10 of A3 is not necessary if L=1. If L>1 then $GOO(X_L)$ and $B(X_L)$ are determined for all orbit representatives $X_L \in SC(\Pi_l)$ since they are needed for A3 application to L=L-1.

A3 algorithm can be applied for determining any $GOO(A(X_{L-1}))$ and $B(X_{L-1})$, L=2, . . . , LK-1 under the following requirements:

R1. For L known are: $\Pi_L$, $SC(\Pi_L)$, a vertex $XF_L \in SC(\Pi_L)$; its index in $SC(\Pi_L, GOO(A(XF_L))$, $B(XF_L)$, $XF_L$ is not similar to any previous vertex in $SC(\Pi_L)$ under $A(X_{L-1})$, $X_L$-selected vertex, bouquets $BFRPO(X_L)$; $GOO(X_{L-1})= GOO(X_L)$ at the start;

R2. $X_L$ is selected in interval index($XF_L$)+1 to the end of $SC(\Pi_L)$, $X_L$ is selected as a first not similar vertex after the current $X_L$ under the current $A(X_{L-1})$. At the start $X_L$ is the vertex with index=index($XF_L$)+1. This means that the vertices of $Orb(XF_L, A(X_{L-1}))$ in $SC(\Pi_L)$ are after $XF_L$. Thus, traversing $SC(\Pi_L)$ we'll traverse $Orb(XF_L,A(X_{L-1}))$– this is the requirement of Theorem 4;

R3. If $X_L$ is similar to some vertex$\in FRPO(X_L)$ then the discovered automorphism $\alpha$ is a generator of $A(X_{L-1})$: gen(A):= gen(A)$\cup \alpha$. Otherwise, each new numeration in $B(X_L)$ (instruction 10) belongs also to $B(X_{L-1})$.

## 5.2. Vsep-e exact algorithm: cases CS1, CS2, CS3 and CS4

Let we have the starting series SFM1: $\Pi_1$, $\Pi_2$,…, $\Pi_{LK}$. We can apply A3 algorithm directly only to the partition $\Pi_{LK-1}$ because for the other partitions the required input variables are not known. For the partition $\Pi_{LK-1}$ we have $B(x_{LK-1}) =\{\Pi_{LK}\}$ and gen($A(x_{LK-1})$)=$\emptyset$, i.e., $|A(x_{LK-1})|=1$ and discrete orbits of $A(x_{LK-1})$. After the application of A3 to $\Pi_{LK-1}$ we have determined correctly $B(x_{LK-2})$ and $GOOGA(A(x_{LK-2}))$. Then, A3 can be applied to $\Pi_{LK-2}$, i.e., a backward move is done from LK-1 to LK-2. Thus, applying A3 to the series $\Pi_{LK-1}$, $\Pi_{LK-2}$,…, $\Pi_2$, $\Pi_1$ we can determine GOO(A). The lowest level to which a backward move has been made we denote by LMIN, i.e., LMIN is the level for which we determine $GOO(A(X_{LMIN-1}))$. In A3 algorithm the process of the backward moves is not included and the instruction 10 is not revealed. All this is taken into account in PART1 (Figure 9) and PART2 ( Figure 11) algorithms called from the Vsep-e algorithm (Figure 8) that determines GOO(A) of the partition-wise stabilizer *A=Aut(G, Π)* given a graph *G(V,E)* and the input partition *Π* on *V*. S3.

Before calling PART1 and PART2 Vsep-e algorithm determines (step S2) the orbits of *A* by an TREE4 heuristic algorithm (see section 7.1) and selects $X_1 \in SC(\Pi_1)$ - a representative of one of the smallest orbit of the vertices in $SC(\Pi_1)$. Experimental tests show with very rare exceptions that if the staring vertex $X_1$ is a representative of one of the smallest orbit of A then the size of the bouquet $B(X_1)$ built by PART1 is the smallest and the running time is minimal. PART1 algorithm (Figure 9) can be considered as an application of A3 algorithm with added the backward moves and revealed instruction 10 – all above requirements are implemented in it. PART1 algorithm determines GOO(A($x_1$)) and B($x_1$) given $\Pi_1$, $SC(\Pi_1)$ and $x_1 \in SC(\Pi_1)$. PART2 algorithm (the second part of Vsep-e algorithm, line S5 in Figure 8) determines GOO(A) given GOO(A($x_1$)) and B($x_1$) obtained from PART1 algorithm.

PART2 algorithm can be considered also as an application of A3 algorithm to the partition $\Pi_1$ with replacing the instruction 10 by determining one derivative numeration $\Pi_{LK}$ of each selected vertex $x \in SC(\Pi_1)$ if x is not similar to $x_1$ under the current A. PART2 algorithm may be also considered as a direct application of theorems 3 and 4. The satisfaction of the requirements of these theorems guarantees .the correctness of PART2 At the start of PART2 GOO(Aut(G)):= GOO(Aut(G,x1)). Each vertex *x* in $SC(\Pi_1)$ that is not similar to x1 under the current Aut(G) is selected (line B2) and a comparison (line B5) of the first numeration $\Pi_{LK}$ derived from the selection x (line B4) with the numerations $\in B(x1)$ is made. If there is an automorphism α between some numeration $\in B(x1)$ and $\Pi_{LK}$ then *α* is a generator for *A* since it unites the orbits of *x* and *x1*. In both cases (existence or nonexistence of *α*) the traversing of $SC(\Pi_1)$ continues until its end. When the traversal of $SC(\Pi_1)$ completes, the generators and the orbits of *A* are determined and the 'orbit-stabilizer' theorem is applied for determining *|A|=|Orb(x1, A)|.| A(x1)|* (line B3).

We'll describe PART1 algorithm considering an intermediate state of ST search tree (Figure 7) being built by the algorithm during its execution. The series of partitions $\Pi_L$, L=2, . . . , LK-1 can be divided into three intervals: the first is from $\Pi_1$ to $\Pi_{LMIN-1}$, the second – from $\Pi_{LMIN}$ to $\Pi_{LP}$ and the third – from $\Pi_{LP+1}$ to $\Pi_{LK-1}$. The search tree is built in a preorder: first visiting the root (a partition $\Pi_L$) and then its subtrees (the partitions $\Pi_{L+1}$ derived from each selected vertex) in a defined order.

Applying A3 algorithm for determining *gen(A($x_L$))* we have:

$gen(A(x_L)) = gen(A(x^1_{L+1})) \cup MG(x^1_{L+1}, x_{L+1}) \cup gen(A(x_{L+1}))$,

where $gen(A(x_L))$ is the generating set of the current stabilizer $A(x_L)$, $gen(A(x^1_{L+1}))$ is the generating set of the stabilizer $A(x^1_{L+1})$, $MG(x^1_{L+1}, x_{L+1})$ is th.e set of the mutual generators of $A(x_L)$ and $gen(A(x_{L+1}))$ is the generating set of the current stabilizer $A(x_{L+1})$. The following conditions hold for the intermediate state of PART1 algorithm execution (Figure 9) shown as a search tree ST on Figure 7:

C1: On the current path of ST tree known are: L, $\Pi_L$, $SC(\Pi_L)$, $X_L$, L=1, 2, . . . ,LK-1, $X_L$ is the current selected vertex in $SC(\Pi_L)$. The current numeration is $H_{LK-1}$=n1;
C2: LMIN, LP, LMIN≤LP, LP≥2 are known;
C3: $X_1$ is the first vertex $\in SC(\Pi_1)$;

C4: Each selected vertex $X_L$ is not similar to any previous vertex in $SC(\Pi_L)$ under $A(X_{L-1})$, L=2,. . . ,LP-1;
C5: Each selected vertex $X_L$ is the first vertex in the $SC(\Pi_L)$ for L=LP+1,. . . ,LK-1; The partitions in this section of the path are a result of the forward move $SFM(X_{LP})$;
C6: Known are: $GOO(X_{LMIN-1})$, $B(X_{LMIN-1})$, $FRPO(X_{LMIN})$;
C7: Known are the bouquets $BFRPO(X_L)$, L=1, 2, . . . , LP;
C8: Known are the orbits and orders $|A(X)|$ of vertices $X \in FRPO(X_L)$ and $X= X_L$, L=LMIN,…,LP, under $A(X_{L-1})$.
C9: On the current path known are the computed orbits $Orb^c(X_l)$ and computed orders $A^C(X_l)$ for L=LMIN+1,…,LK-1 under the current $A(X_{lmin-1})$. For L=LMIN we have $Orb^c(X_l)= Orb(X_l)$ and $A^C(X_l)= A(X_l)$ under the current $A(X_{lmin-1})$ since all generators found so far have the same fixed points with $A(X_{lmin-1})$.
The action that follows the above state is a comparison of n1 numeration with the numerations of $BFRPO(X_{LP})$ for discovering a new generator of $A(X_{lmin-1})$.
All above conditions can be considered as an invariant for correctness of the loop C1 of PART1algorithm.
Let's now describe PART1 algorithm. It calls SFM1 (Figure 10) and COMP (Figure 13) algorithm. At the start all of the searched variables are not known and for each partition $\Pi_L$, L=2, . . . , LK-1 we select the first vertex $X_L \in SC(\Pi_L)$ and obtain the partition $\Pi_{L+1}$, i.e. the only action we do is a forward move (line I1) until a discrete $\Pi_{LK}$ is obtained. Thus, the conditions R1 to R3 hold for only for $\Pi_{LK-1}$. Let's now consider the above requirements R1 to R3 for determining $GOO(A(X_{LP-1}))=GOO(A(x_1, , . . . , x_{LP-1}))$ and the bouquet $B(X_{LP-1})=B(x_1, , . . . , x_{LP-1})$ given the partitions $\Pi, \Pi_1, . . . , \Pi_{LP}$. For $\Pi_{LP}=\Pi(x_1, , . . . , x_{LP-1})$ also $SC(\Pi_{LP})$ and the selected vertex $X_{LP}$ are known. Besides, the requirements hold for the position of the current vertex $X_{LP}$ in $SC(\Pi_{LP})$. By the loop C1 (lines I1- I12) in PART1 algorithm each selected cell $SC(\Pi_{LP})$ is visited and the following four basic steps are performed:
A1 {*Selection*}: The selection of a vertex $X_{LP}$ in $SC(\Pi_{LP})$ (line I3) is made starting from the position next to the current $X_{LP}$. The vertex $X_{LP}$ should not be similar to any previous vertex in $SC(\Pi_{LP})$ under the current $A(X_{LMIN-1})$. For each level L the position $i(X_L)$ of the selected vertex $X_L$ is stored and when a backward move to this level is performed then the selection of a new vertex starts from the next position, i.e. $i(X_L)+1$. If there is no selected vertex in $SC(\Pi_{LP})$, i.e. the $SC(\Pi_{LP})$ has been traversed then, a backward move follows (step A4). If there is a selected vertex $X_{LP}$ in $SC(\Pi_{LP})$ then, step A2 follows.
A2 {*Series of forward moves*}: A series of forward moves SFM1 is performed determining the partitions $\Pi_L=\Pi_{LP+1}, . . . , \Pi_{LK}$ with discrete $\Pi_{LK}$ (line I5). In each of these partitions the selected vertex $X_L$ is the first vertex in $SC(\Pi_L)$. This way the requirements hold for the orbit $Orb(X_L, A(X_{L-1}))$.
A3 {*Comparison*}: A check if there is a new automorphism $\alpha$ that not belong to the current subgroup of $A(x_{LP-1})$ and maps the vertex X to any vertex from FRPO(X) is made, i.e. if $\alpha$ belongs to the next subgroup of $A(x_{LP-1})$. This check is made by comparing $\Pi_{LK}$ with BFRPO(X) (line I6, COMP algorithm).
A4 {*Backward move*}: After the traversal of $SC(\Pi_{LP})$ is completed then $GOO(A(x_{LP-1}))$ and $B(x_{LP-1})$ are determined and a backward move

LP:=LP-1 is made. Stop follows if LP=1. Otherwise, a selection of a new
vertex in SC($\Pi_{LP}$) is made applying the step A1 to it.
If there is an automorphism α then, it is a generator: gen(A($X_{LP-1}$)) := gen(A($X_{LP-1}$))∪{α}
and the orbits and the order of A($X_{LP-1}$) are recomputed. It is a generator also for A($X_1$):
gen(A($X_1$)):=gen(A($X_1$))∪{α} and the orbits and the order of A($X_1$) are recomputed. If
there is no α mapping x to a vertex∈FRPO(X) then a move back to LK-1 follows. This
way the building of the tree ST(x) starts from LK-1 performing the step A1 to SC($\Pi_{LK-1}$).
ST(x) is necessary since it determines the bouquet B(x) that belongs to B($X_{LP-1}$). After
ST(x) has been built we continue with a selection of a new vertex in SC(LP) applying the
step A1 to SC($\Pi_{LP}$). The search tree is built in preorder traversal: first visiting the root ($\Pi_{LP}$
partition) and then its subtrees (the partitions $\Pi_{LP+1}$ derived from each selected vertex) in a
defined order.
For the selected vertex $X_{LP}$ (line I3 – the start of ST($X_{LP}$) building) by SFM1 (line I5) is
built the first (leftmost) tree ST($X_{LP+1}$), ST($X_{LP+2}$), . . . , ST($X_{LK-1}$) for each previous
subtree. Each of these subtrees is built in backward ord..er. When the subtree ST($X_{LP+1}$) has
been built then the building of the subtree for the next selected vertex $X_{LP+1}$ starts ($X_{LP+1}$
should hold the requirements). When there is no selected vertex $X_{LP+1}$ then a backward
move LP+1 to LP is made – this means that the ST($X_{LP}$) is built. If
LP=1 the algorithm stops.

### 5.2.1. Cases CS1 and CS3

**Figure 7. Search tree of VSEP-e**

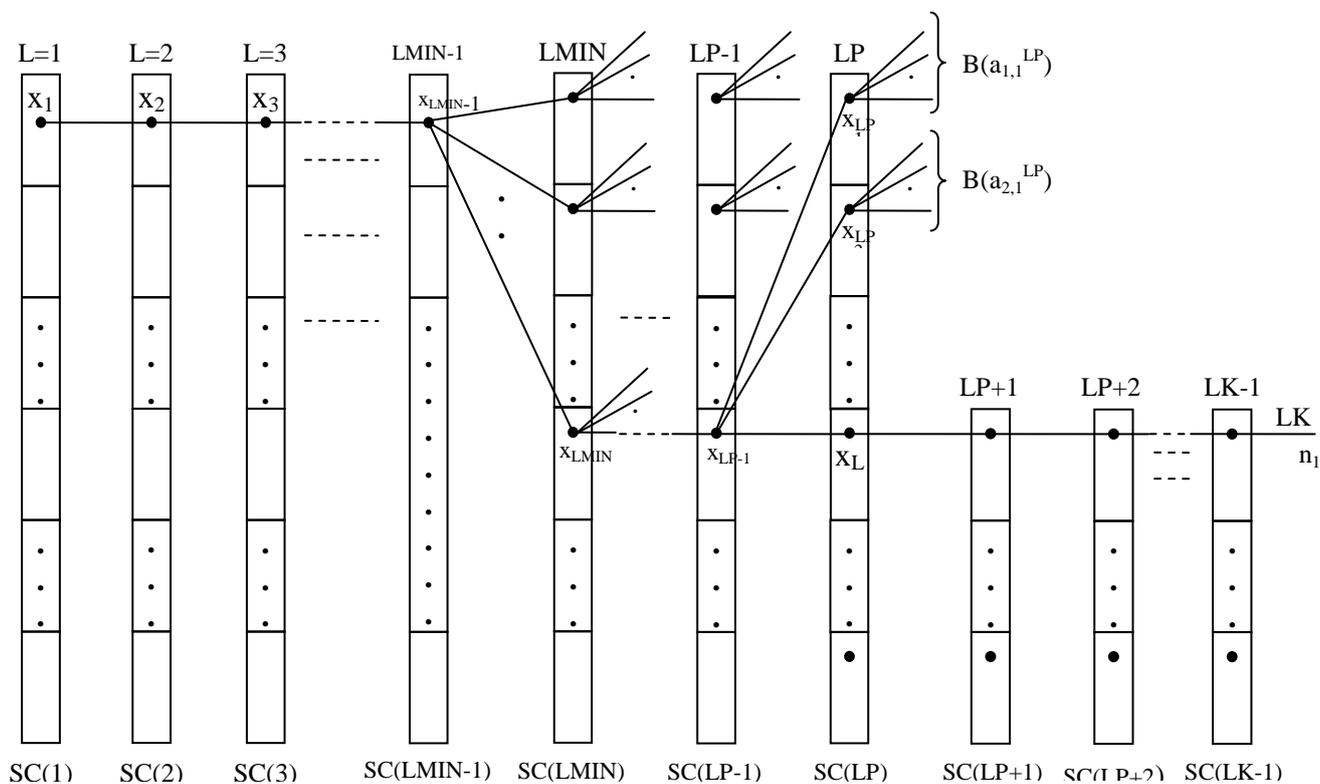

Let us consider the cases when the numeration $.\Pi_{LK}$ (Figure 7) does not form an automorphism with any numeration $\in BFRPO(\Pi_{LP})$. Knowing that $|A(X_{LP-1})|$ is correct and supposing that each orbit $Orb(X_L, A(X_{L-1}))$ for $L=LP, \ldots, LK-1$ is also correct, and applying the Theorem 'O-S' we obtain

$|A(X_{LP-1})|/|Orb(X_{LP}, A(X_{LP-1}))|/|Orb(X_{LP+1}, A(X_{LP}))|/ \ldots /|Orb(X_{LK-1},A(X_{LK-2}))| =$
$|A_{LK-1}| = 1.$  (5.2.1)

However, the orbits in (5.2.1) are unknown and consequently we cannot use it. Instead, the sets $W=COrb(X_L, A(X_{LP-1}))$ of the vertices in $SC(\Pi_L)$ similar to $X_L$ under $A(X_{LP-1})$ are known, i.e.

$W=COrb(X_L, A(X_{LP-1}))= Orb(X_L, A(X_{LP-1}))\cap SC(\Pi_L).$  (5.2.2)

We call the set W *'computed orbit'* to distinguish it from the real orbit $U=Orb(X_L, A(X_L))$. The following relation holds $U=Orb(x_L,A(x_{L-1}))\subseteq COrb(x_L,A(x_{LP-1}))=W,$  (5.2.3)
since $A(x_{L-1})) \subseteq A(x_{LP-1})$.

Considering (5.2.3) and applying the Theorem 'O-S' for the computed orbits we obtain
$|A(X_{LP-1})|/|COrb(X_{LP},A(X_{LP-1}))|/|COrb(X_{LP+1},A(X_{LP-1}))|/ \ldots / |COrb(X_{LK-1},A(X_{LP-1}))| =$
$|A_{LK-1}| \leq 1.$  (5.2.4)

The computation on (5.2.4) is performed by SFM1 algorithm (Figure 10) called from the instruction I5 of PART1. If the sign in (5.2.4) is = (i.e., the real and the computed orbit are equal), then we call the orbit $Orb(x_L, A(x_{L-1}))$ *separated*. If the sign in (5.2.4) is <, then, obviously, $W=U\cup R$, $|R|>1$, i.e., W (and R) include vertices that belong to $SC(\Pi_L)$ but are not similar to $X_L$ under $A(X_{L-1})$. It can be proved that when W contains some vertex, then it contains the whole orbit of this vertex under $A(X_{L-1})$.

Thus, W can be considered as an union of orbits of $.A(X_{L-1})$. We call this case ***non-separation of orbits (denoted by NSO)*** and the orbit U is called ***non-separated (non-partitioned).*** Since the representatives of the orbits belonging to R are not known we cannot select them during the traversal of the $SC(\Pi_L)$. Thus, the search tree of such a

---

*Input*: Graph *G(V,E)* and a partition *Π* on V
*Output*: Generators, orbits and order of the *partition-wise stabilizer A=Aut(G, Π),* denoted *GOO(A)*

S1: Initialization: $\Pi_1:=RP(\Pi, NCL);\{NCL$-the number of the cells in $\Pi_1\}$
S2: **if** NCL=n {$\Pi_1$ is discrete} **then** message'Trivial group'); **return end if**;
S3: CSLCh {call cell selector chooser-finds the cell selector};
S4: Determine $SC(\Pi_1)$ and $x_1:=$first vertex in $SC(\Pi_1)$;
S5: Use a filter that defines whether or not to call the following procedures that:
  a) Give a better partition $\Pi_1$ obtained by S code sorting of the vertices in $\Pi_1$. Determine a new $SC(\Pi_1)$ and $x_1:=$first vertex in $SC(\Pi_1)$;
  b) Determine: (i) the generators and orbits of *A* by TREE4 heuristic algorithm, starting from discrete orbits; (ii) $SC(\Pi_1)$; $X_1:=$a representative of the smallest orbit of the vertices in $SC(\Pi_1)$; Put the vertex $x_1$ on the first position in $SC(\Pi_1)$;
S6: PART1: Determine $GOO(A(x_1))$, $B(x_1)$ given $\Pi_1$, $SC(\Pi_1)$, $x_1\in SC(\Pi_1)$
S7: PART2: Determine $GOO(A)$ given $GOO(A(x_1))$, $B(x_1)$

**Figure 8. Vsep-e algorithm**

vertex cannot be built and its bouquet will not be determined. This is an unallowable error since these bouquets belong to $B(X_{L-1})$, $B(X_{L-2})$, . . . , $B(X_1)$ and they are needed (as we know from A3 algorithm) for determining
$GOO(A(X_{L-2}))$, $GOO(A(X_{L-3}))$, . . . , $GOO(A(X_1))$. If there is only one partition with non-separated orbit, then the sign in (4.2.1.4) is <, since the length of the computed orbit is greater or equal to the length of the real orbit – this is the condition to detect the presence of NSO. We call this case CS3. If there is no NSO in any partition, i.e. each computed orbit is equal to the real one, then the sign in (4.2.1.4) is = and the case is denoted by CS1. If the case is CS1, then at the exit of COMP algorithm (line I6 in PART1) the invariant for correctness holds for LP=LK-1: only LP is changed, LMIN remains the same. When the case is CS3 we lose $GOO(X_{LMIN-1})$ determined so far and the algorithm continues with a new start point: LP=LK-1, LMIN=LP, $X_{LMIN}$= $X_{LK-1}$ is the first vertex in SC(LK-1) and $A(X_{LMIN-1})$={I}, i.e., each vertex is put into a separate orbit. Obviously, the invariant for correctness holds for CS3 case.

---

**Input**: $\Pi_1$, $SC(\Pi_1)$, $x_1$
**Output**: $GOO(A(x_1))$, $B(x_1)$
I1:SFM1:Determine $\Pi_L$,$|A(X_{L-1})|$=1,SC(L),L=2, . . . ,LK given ,$\Pi_1$,$SC(\Pi_1)$,$X_1$;
    LP:=LK-1; LMIN:=LP; gen($A(x_1)$):=∅; $|A(x_1)|$:=1; $B(X_{LK-1})$:={$\Pi_{LK}$};
I2: **do** *{loop C1: Visit vertices in $SC(\Pi_{LP})$ }*
I3:   Select $X_{LP}$ in SC(LP);
I4:   **if** $X_{LP}$≠0 **then** *{forward move}*
I5:     SFM1: Determine $\Pi_L$, $|A(X_{L-1})|$, SC(L),L=LP+1, . . . ,LK given LP, $\Pi_{LP}$, $SC(\Pi_{LP})$, $X_{LP}$;
I6:     COMP: Compare $\Pi_{LK}$ with the numerations∈$BFRPO(X_{LP})$ for determining an automorphism α∈$A(X_{LP-1})$. Determine $GOO(A(X_{LMIN-1}))$, LP;
I7:   **else** *{backward move: SC(LP) has been traversed}*
I8:     LP:=LP-1;
I9:     **if** LP<LMIN **then** LMIN:=LP; $|Aut(X_{LMIN})|$:= $|Aut(X_{LMIN+1})|$ **end if**;
I10:   **if** LP=1 **then return end if**;
I11:   **end if**
I12:**end do**;*{ loop C1}*

**Figure 9. PART1 algorithm (instruction S6 of Vsep-e algorithm (Figure 8))**

---

**Input:** L=LP, $\Pi_L$, $SC(\Pi_L)$, $X_L$; **Output:** $\Pi_L$, $|A(X_{L-1})|$, L=LP+1,… ,LK
1. **do**
2.   L:=L+1; $\Pi_L$:=RP($\Pi_{L-1}$,$X_{L-1}$,BRCL);
3.   **if** BRCL=n **then return end if**
4.   determine SC(L); $X_L$:=first vertex in SC(L);
5.   $|A(X_L)|$:= $|A(X_{L-1})|$/$|Orb(X_L, A(X_{LMIN-1})\cap SC(\Pi_L)|$
5. **end do**

**Figure 10. SFM1 algorithm (instruction I5 of PART1 algorithm (Figure 9)**

The operations in the case CS3 may be considered as an error correction of the incorrect orbits of some $A(X_{L-1})$ determined by the moment since the algorithm interrupts its current execution and starts from the new start point for finding the correct orbits of $A(X_{L-1})$ and the bouquet $B(X_{L-1})$.

### 3.2.1. Cases CS2 and CS4

Let's consider the cases when there is an automorphism α mapping $X_{LP}$ to some vertex $U_{LP} \in FRPO(X_{LP})$, i.e. the numeration $\Pi_{LK}$ forms an automorphism α with some numeration $\in B(U_{LP})$, $X_{LP} = \alpha(U_{LP})$.

---

**Input**: L=1, $\Pi_1$, $SC(\Pi_1)$, $x_1$, $GOO(A(x_1))$, $B(x_1)$
**Outut**: $GOO(A)$

B1: **do**
B2:   Selecte next vertex X in $SC(\Pi_1)$;
B3:   **if** X=0 **then** $|A|=|Orb(x_1, A|.|A(x_1)|$; **return end if;**
B4:   SFM1A: Determine $\Pi_{LK}$ given L=1, $\Pi_1$, $SC(\Pi_1)$, $X \in SC(\Pi_1)$
B5:   COMP: Compare $\Pi_{LK}$ with the numerations $\in$ bouquet $B(x_1)$ for an automorphism α. If there is α then $gen(A):=gen(A) \cup \alpha$ and recompute the orbits of A;
B6: **end do**;

---

**Figure 11.** PART2 algorithm (instruction S5 of Vsep-e algorithm (**Figure 8**))

---

**Input:** L=1, $\Pi_1$, $SC(\Pi_1)$, $X \in SC(\Pi_1)$
**Output:** $\Pi_{LK}$

1: **do**
2:   L=L+1; $\Pi_L = IR(\Pi_{L-1}, X_{L-1})$;
3:   **if** NCL=n **then return end if;**
4:   **else** determine SC(L); $X_L$=first vertex in SC(L)
5: **end do**

---

**Figure 12.** SFM1A algorithm (instruction B4 of PART2 algorithm (    **Figure 11**))

Then, there is a possibility of NSO for some orbits of the vertices of the target cells of the current path for the levels LMIN+1 ≤ L ≤ LP: we denote by CS2 the case when there is no NSO and by CS4 the case when there is at least one case of NSO in this interval. The automorphism α is a generator of $A(X_{LMIN-1})$ since $A(X_{LP-1}) \subseteq A(X_{LMIN-1})$: $gen(A(X_{LMIN-1})) := gen(A(X_{LMIN-1})) \cup \{\alpha\}$. Thus, the $Orb(A(X_{LMIN}))$ and $|A(X_{LMIN})|$ are changed (line 11) and we denote by $^{+\alpha}$ any variable with changed value. Before α each orbit and each order of the current stabilizers are correct: $|A(Z_L)| = |A(X_{L-1})|/|Orb(Z_L, A(X_{L-1}))|$ for each representative of orbit $Z_L \in SC(L)$, L=LMIN, . . . , LP.
The following actions are:

a) For the interval L=LMIN to LP (loop: lines 11 to 24) of the current ST path the stabilizers of the selected vertices $X_L$ before and after $\alpha$ are determined (line 13) ;

| |
|---|
| **Input:** $\Pi_L$, $SC(\Pi_L)$, $X_L$, $\|A(X_{L-1})\|$, L=2,3, . . . , LK; LP, $BFRPO(X_{LP})$, LMIN, $GOO(A(X_{LMIN-1}))$ |
| **Output:** $GOO(A(X_{LMIN-1}))$, LP |
| 1: Compare $\Pi_{LK}$ with the numerations $\in BFRPO(X_{LP})$ |
| 2: **if** $\Pi_{LK}$ is not equivalent to any numeration $\in BFRPO(X_{LP})$ |
| 3:   **then** {CS1 or CS3} |
| 4:     **if** $\|A(X_{LK-1})\|=1$ |
| 5:       **then** {CS1} |
| 6:         LP=LK-1; $B(X_{LP})=\{\Pi_{LK}\}$; **if** LP<LMIN **then** LMIN=LP |
| 7:       **else** {CS3} |
| 8:         LP=LK-1; $B(X_{LP})=\{\Pi_{LK}\}$; LMIN=LP; $A(X_{LMIN-1})=\{I\}$; $X_{LMIN}$:=the the first vertex in SC(LMIN); $\|Aut(X_{LMIN})\|$:=1 |
| 9:     **end if** |
| 10:   **else**{CS2 or CS4: there is an automorphism $\alpha$, $X_{LP}=\alpha(U_{LP})$, between $\Pi_{LK}$ and some numeration $\in BFRPO(X_{LP})$ derivative of the vertex $U_{LP} \in FRPO(X_{LP})$} |
| 11:     Determine $GOO(A^{+\alpha}(X_{LMIN-1}))$: $gen(A^{+\alpha}(X_{LMIN-1}))=A(X_{LMIN-1}) \cup \{\alpha\}$; determine orbits of $A^{+\alpha}(X_{LMIN-1})$ and $\|A^{+\alpha}(X_{LMIN-1})\|=\|A(X_{LMIN})\|\|Orb(X_{LMIN}, A^{+\alpha}(X_{LMIN-1}))\|$ |
| 12:   **for** L=LMIN, LMIN+1, . . . , LP **do** |
| 13:     $\|A(X_L)\|=\|A(X_{L-1})\|/\|Orb(X_L, A(X_{L-1}))\|$; $\|A^{+\alpha}(X_L)\|=\|A^{+\alpha}(X_{L-1})\|/\|COrb(X_L, A^{+\alpha}(X_{LMIN-1}))\|$; |
| 14:     **if** L>LMIN **then** |
| 15:       **for** each vertex $Z_L \in FRPO(X_L)$ **do**{check if $\|A(Z_L)\|$ has changed after $\alpha$} |
| 16:         **If** $\|A^{+\alpha}(Z_L)\|=\|A(X_{L-1})\|/\|COrb(Z_L, A(X_{LMIN-1}))\| < \|A(Z_L)\|=\|A(X_{L-1})\|/\|Orb(Z_L, A(X_{L-1}))\|$ |
| **17:**         **then** CS4=true |
| **18:**         **end if** |
| **19:**         **if** L=LP and index($Z_L$)= index($Z_{Lp}$) |
| 20:         **then** IULP:= index($U_{LP}$); RSTBULP:=$\|A(U_{LP})\|$ |
| 21:         **end if** |
| 22:       **end for** {loop for from line 15} |
| 23:     **end if** |
| 24:   **end for** {loop for from line 12} |
| 25:   **if** (not CS4) **then** return **end if** |
| 26:   **if** RSTBULP=1 |
| 27:   **then** LMIN:=LP;L:=LP; $gen(A(X_{LMIN-1})):=\{\alpha\}$; orbits of $A(X_{LMIN-1})$ :=cycles of $\alpha$; $\|A(X_{LMIN-1})\|:=\|Orb(U_{LP}, A(X_{LMIN-1}))\|$; $XF_{LMIN}:=U_{LP}$; $\|A(XF_{LMIN-1})\|:=1$; the execution continues by starting the selection with a vertex with index($U_{LP}$)+1 |
| 28:   **else** $SFM(U_{LP})$**;L:=LK-1;LP:-L;LMIN:=L;**$A(X_{LMIN}):=\{I\}$;the next selected vertex is the first vertex in SC(LP) |
| 29:   **end if** |
| 30:**end if** |

**Figure 13. . COMP algorithm (instruction I6 of PART1 algorithm (Figure** 9))

b) For each representative $Z_L \in FRPO(X_L)$, L>LMIN a check if $|A(Z_L)|$ has changed after $\alpha$ (line 16) is performed. This check is excluded for L=LMIN since all orbits and
c) stabilizers are correct – all so far found generators have the same fixed points with $A(X_{LMIN-1})$; If there is a change of $A(Z_L)|$ for some L then CS4=true. Also, the index($Z_{LP}$) and $|A(U_{LP})|$ are stored;
d) If CS4=false (line 25), i.e. the case is CS2 then an exit from COMP follows, no changes of the variables computed so far are made and the next selection starts from the current selected vertex in SC (LP).
e) In CS4 $GOO(X_{LMIN-1})$ found so far are lost and the algorithm continue from a new start point as in CS3 case. The invariant for correctness holds for the new state of the ST tree. The main requirement that hold is that the selected vertices $X_L$ of the current path ($U_{LP}$ for L=P) are not similar to the previous vertices in SC(L) under $A(X_{L-1})$. If the case is CS4 (lines 26 to 30) then there are possible two subcases: i) $|A(U_{LP})|=1$ and ii) $|A(U_{LP})|>1$. If $|A(U_{LP})|=1$ then LMIN:=LP, L:=LP and a new $GOO(X_{LMIN-1})$ determined (line 17) are determined – the new start point is index($U_{LP}$) in SC(LP). The case when $A(U_{LP})|>1$ (line 28) means that there are generators we do not know – that's why we do a forward move FM(index($U_{LP}$)) to come to a new start point: L=LK-1;LP=L, LMIN=L, $A(X_{LMIN}):=\{I\}$;the next selected vertex is the first vertex in SC(LP). From registered CS4 cases in experiments on benchmark graphs there are only cases with $|A(U_{LP})|=1$.
There is another way for CS2|CS4 check (not shown in COMP procedure): the check for difference of the new and old stabilizer is made only for the selected vertices $X_L$ of the path. The experiments show that both ways work correctly.

The experiments on the benchmark graphs in [19] show that CS4 case occurs only for the graphs B52 (Mathon doubling of b25-1 graph [18]), latin-16 and 24, and had-96.

## 5.3. Examples

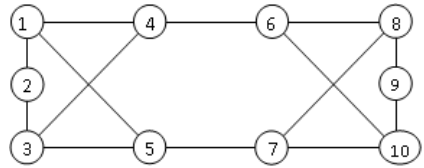

**Figure 14. G10 graph**

### 5.3.1. Simple example

Let us consider the search tree traversal in Figure 15 (in preorder) for the graph of Figure 14. Starting from the partition $\Pi_1 = \Pi() = RP(\Pi_u) = |2,9|4,5,6,7|1,3,8,10|$ we do a series of forward moves SFM: $\Pi_2 = \Pi(4) = |2|9|4|7|5|6|8,10|1,3|$, $\Pi_3=\Pi(4,8) = |2|9|4|7|5|6|8|10|1,3|$, $\Pi_4=\Pi(4,8,1)=|2|9|4|7|5|6|8|10|1|3|$ - the first numeration $n_1$. The selected cell $SC(\Pi_L)$ in this SFM is the cell with the largest number of adjacency cells, the selected vertex is always the first in $SC(\Pi_L)$ and the order of the stabilizers $A^{(0)}, A^{(1)}, A^{(2)},\ldots, A^{(L)}$, L=1,2,…,LK-1 is 1. Then, a back move follows, L=L-1, L=3, vertex 3 is selected and $\Pi_4 = \Pi(4,8,1)= |2|9|4|7|5|8|10|3|1|$ - $n_2$, $n_2 = \alpha_1(n_1)$, $\alpha_1=(1,3)$. Then, again back move to L=3, where there is no selected vertex in SC(3), new back move to L=2, $|A(4,8)|=|Orb(1)|.|A(4,8,1)|=2,1=2$ and

- 29 -

selected vertex in SC(2) is 10, $\Pi_3=\Pi(4,10) = |2|9|4|7|5|6|10|8|1,3|$, $\Pi4_3 = \Pi(4,10,1) = |2|9|4|7|5|6|10|8|1|3|$ - $n_3$, $n_3 =\alpha_2(n_1)$, $\alpha_2=(8,10)$. At this point, there are no selected vertices in SC(3) and SC(2), that's why back moves to L=3,2,1 are made and $|A(4)| = |Orb(8)|.|A(4,8)|=2.2=4$. Here PART1 ends and PART2 starts generating two SFM, for vertices 5, 6 $\in$ SC(1). The first SFM is: $\Pi(5)= |2|9|5|6|4|7|6|8,10|1,3|$, $\Pi(5,8)=|2|9|5|6|4|7|6|8|10|1,3|$, $\Pi(5,8,1)= |2|9|5|6|4|7|6|8|10|1|3|$ - $n_4$, $n_4 =\alpha_3(n_1)$, $\alpha_3=(4,5).(6,7)$. The second SFM is: $\Pi(6)= |9|2|6|7|5|4|1,3|8,10|$, $\Pi(6,1)= |9|2|6|7|5|4|1|3|8,10|$, $\Pi(6,1,8)= |9|2|6|7|5|4|1|3|8|10|$ - $n_5$, $n_5 =\alpha_4(n_1)$, $\alpha_4=(1,8)(2,9)(3,10)(4,6).(5,7)$. Orbits of A are: (2,9)(1,3,8,10)(4,5,6,7) and $|A|=|Orb(4)|.A(4)=4.4=16$.

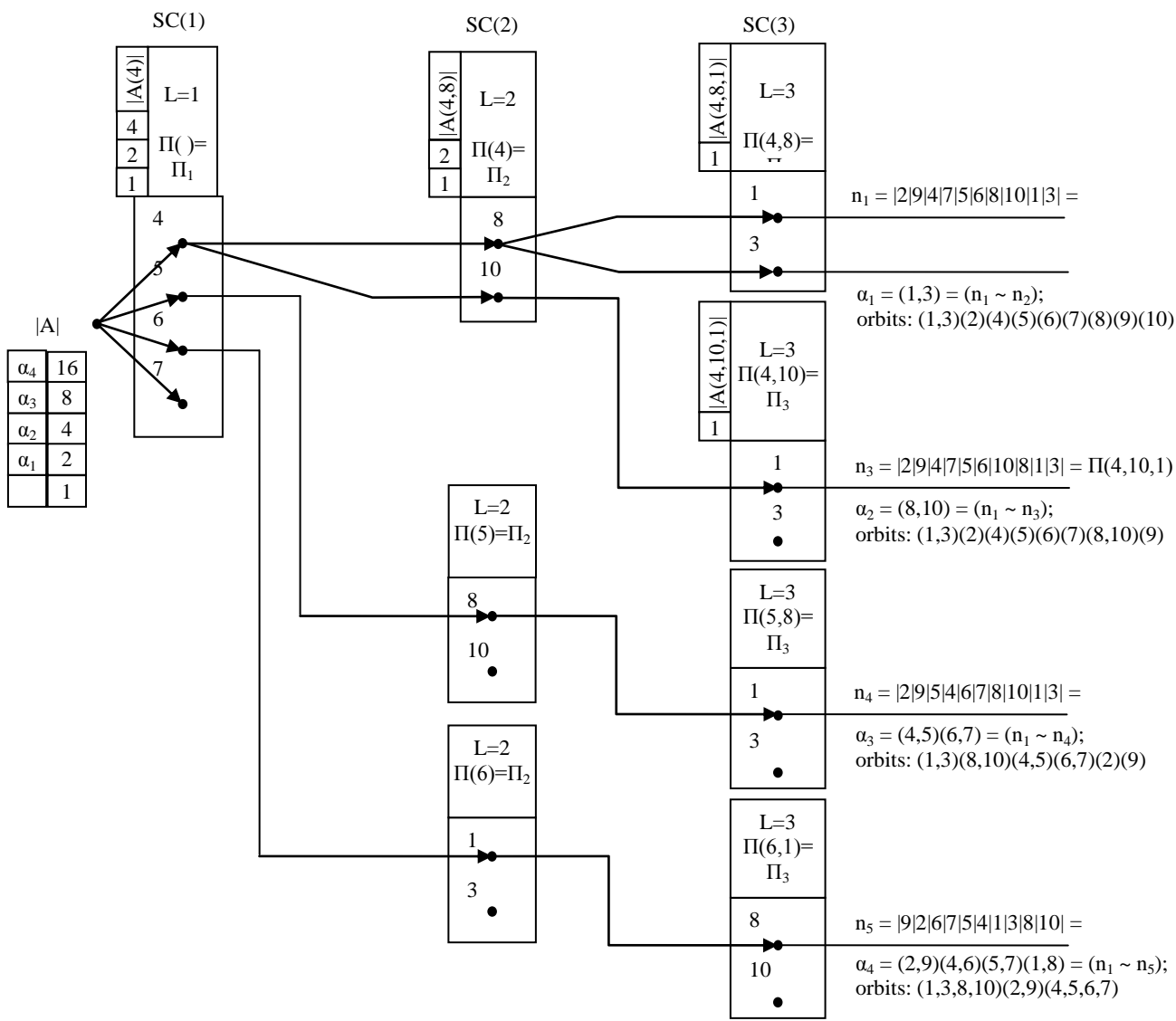

**Figure 15. Search tree for the graph of Fig, 14**

*5.3.2. Example with all cases (CS1, CS2, CS3, CS4) (Figure 16 )*

 The example is for graph G=B52 (Mathon [29]), regular graph, n=52, degree=25, |Aut(G)|=12, orbit lengths: 2*2+4*6+2*12; orbits:
(4,30)(24,50) (21,37, 12,38,11,47) (17,36,16,42,10,43) (8,44,39,18,13,34) (26,35,28,2,9,52) (29,25,22,15,20,33,7,46,41,48,3,51) (40, 6,49,45,5,1,27,31,19,23,32,14).
We show in Figure 16 only the subtrees of the selections (4,11), (4,13), (4,14) and (4,24) of the search tree. The first selected vertex in Π1=|1,2, . . . ,52| is the vertex 4 since it is from one of the smallest orbits – the orbits are found by the heuristic algorithm. We start the consideration from the selections (4,11,51) – this is the numeration n52: it is not equivalent to any numeration from B(4,2), B(4,3), B(4,10). Before these selections there were determined the bouquets of the representatives B(4,2), |B(4,2)|=9, B(4,3), |B(4,3)|=37, B(4,10), |B(4,10)|=15, i.e., totally 51 nonequivalent numerations. These bouquets are derivatives of FRPO of the set {2, 3, 7, 9, 10} – these vertices precede the selected vertex 11 in SC(Π2), Π2=Π(4). There are no bouquets for the vertices 7 and 9 since they are similar to previous vertices in SC(Π2) under A(4): 7-3, 9-2. We have LP=LMIN=2 at the selections (4,11,51). Before these selections there are found 4 generators of A(4), its order, orbits and some stabilizers. At the selections (4,11,51) the case CS3 has been discovered. That's why the selected vertex 51 in SC(Π3), Π3=Π(4, 11) becomes a new starting start point: all information about the stabilizer A(4) obtained so far is lost and
A(4)=A(4,11)=A(4,11,51)={I}, LP=LMIN=3, XFLMIN=51, |A(XFLMIN)|=1. The next selected vertex in SC(Π3) is 7. The numeration Π(4, 11, 7) is equivalent to numeration n52.
Thus a new generator $α_5$ for A(4,11) and A(4) is found (this is case CS2), $α_5$= (1,6)(2,52)(3,41)(4)(5,45)(7,51)(8,39)(9)(10,43)(11)(12,21)(13,34)(14,23)(15,29)(16)(17,36)(18)(19,31)(20,22)(24)(25,33)(26,28)(27,32)(30)(35)(37)(38,47)(40,49)(42)(44)(46,48)(50). Then, we compute
|A($X_{LMIN-1}$)|=|A($XF_{LMIN}$)||Orb($XF_{LMIN}$, A($X_{LMIN}$))|, i.e., |A(4,11)|=|Orb(51, A(4,11))|.|A(4,11,51)|=2*1=2. The next selected vertex in SC(Π3) is 9 - the partition $Π_4$=Π(4, 11,9) is not discrete: |A(4, 11,9)|= |A(4, 11)|/Orb(9, A(4, 11))|=2/|{9}|=2/1=2. Then, the next selected vertex in SC($Π_4$) is 46 and Π5=Π(4, 11,9,46) is discrete (numeration n53). The numeration n53 is not equivalent to the numeration n52 and |A(4, 11,9,46)|= |A(4, 11,9)|/Orb(46, A(4, 11,9))|=2/|{46,48}|=2/2=1 – this is CS1. The vertex 48 is not selected in SC($Π_4$) since it is similar to the vertex 46 under A(4, 11,9). Then, a backward move to L=3 and a selection of the vertex 10 are made. The partition n54 is discrete and not equivalent to any numeration in B(4,11), |A(4,11,10)|= |A(4,11)|/Orb(10, A(4,11))|=2/2=1 (the case is CS1). The vertex 43 in SC(Π3) is not selected because it's similar to the vertex 10 under A(4, 11). The next selected vertex in SC(Π3) is 48- the partition $Π_4$=Π(4,11,48) is not discrete, so we do forward move to L=4 and choose vertex 27 in SC($Π_4$). The numeration $Π_5$=Π(4,11,48,27)=n55 is not equivalent to any numeration in B(4,11): |A(4,11,48,27)|=|A(4,11)|/Orb(48,A(4, 11))|/Orb(27, A(4,11,48))| = 2/|{46,48}|/|{27,32}| =2/2/2=0.5<1. This is case CS3. So, the vertex 27 in SC(Π(4,11,48)) becomes a new starting start point: all information about the stabilizer A(4,11) is lost, A(4)=A(4,11)=A(4,11,48)=A(4,11,48,27)={I}, LP=LMIN=4, $XF_{LMIN}$=27, |A($XF_{LMIN}$)|=

**Figure 16. The search tree ST for the graph B52**

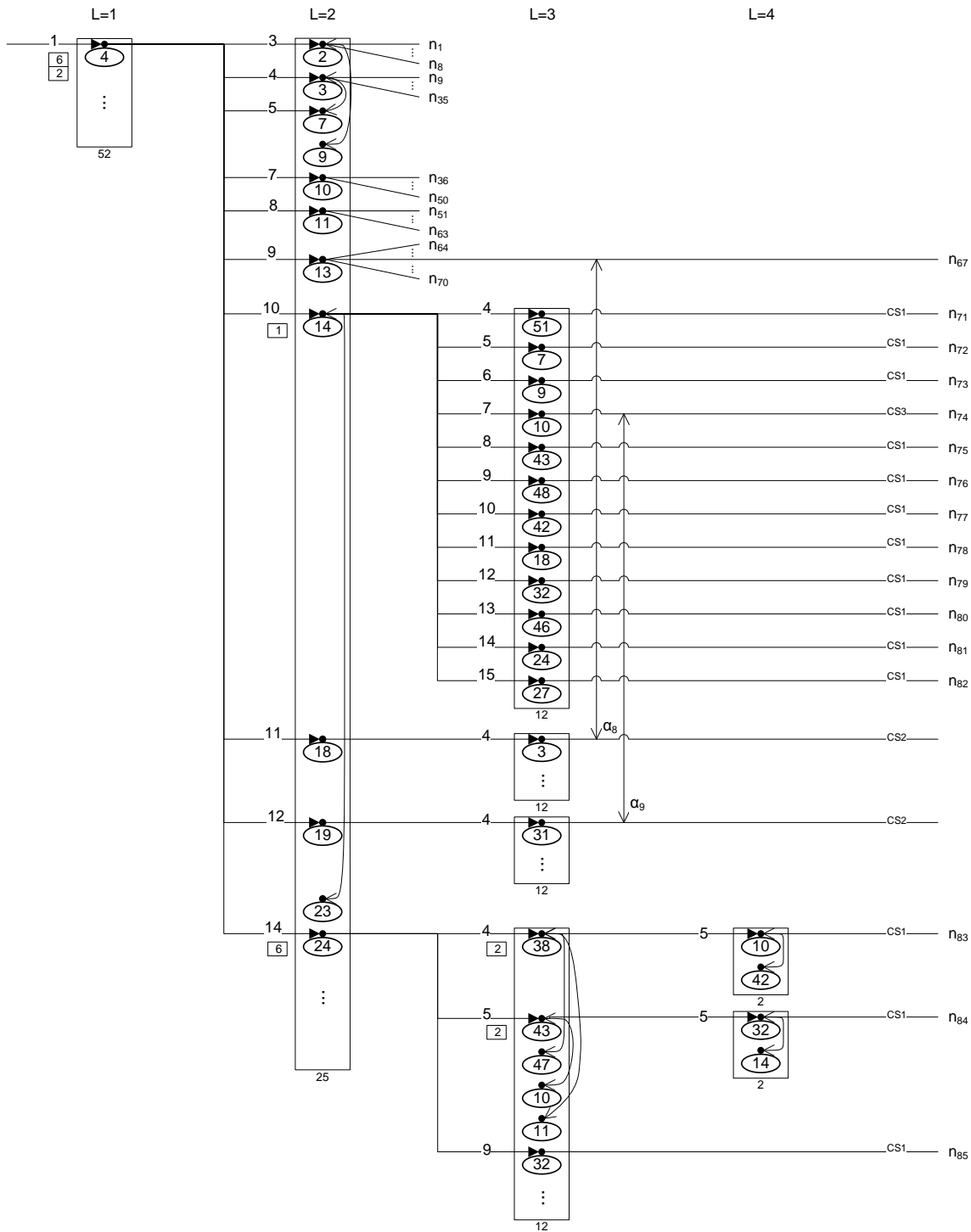

**Figure 16A. The search tree ST for the graph B52 (continued)**

$|A(4,11,48,27)|=1$. We omit the description of the next selections in $SC(\Pi_3)= SC(\Pi(4,11))$. We only mention the occurrence of CS2: the generator $\alpha_6=\alpha_5$ that leads to $gen(A(4,11))= \{\alpha_6\}$, $|A(4,11)|=2$ and the orbits of $A(4,11)$ equal to the cycles of $A(4,11)$. After the $SC(\Pi(4,11))$ has been traversed a backward move to L=2 follows: LP=LMIN=2, $B(4,11)=\{n52,n52,...,n64\}$, $B(4)=\{n1,n2,...,n63\}$), $XF_{LMIN}=11$, $|A(XF_{LMIN})|= |A(4,11)|=2$. The next selected vertex in $SC(\Pi_2)$ is 13: the partition $\Pi_3=\Pi(4,13)$ is not discrete, we do forward move to L=3 and choose the vertex 3 in $SC(\Pi_3)$. The numeration $\Pi_4=\Pi(4,13,3)$

(n64) is not equivalent to any numeration in B(4). The order of its stabilizer is
|A(4,13,3)|=|A(4)|/|Orb(13, A(4))|/|Orb(3,A(4,13))| =2/|{13,34}|/|{3}|
=2/2/1=1 (CS1). After the selection 47 in SC($\Pi_3$) and 31 in SC($\Pi_4$) we obtain the
numeration n65=$\Pi_5$=$\Pi$(4,13,47,31) that is not equivalent to n65 and
|A(4,13,47,31)|=|A(4)|/|Orb(13,A(4))|/|Orb(47,A(4,13))|/|Orb(31,A(4,13,47))|=2/|{13,34}|/|
{47}|/|{31}|=2/2/1/1=1 (CS1). After the selection 14 in SC($\Pi_4$) we have the numeration
$\Pi$(4,13,47,14), LP=4, LMIN=2. Numeration $\Pi$(4,13,47,14) is equivalent to previous one
n65. Thus, a new generator $\alpha_7$ = (1,49)
(2,9)(3,46)(4)(5,40)(6,45)(7,41)(8,44)(10)(11,38)(12,37)(13)(14,31)(15,33)(16,17)(18,34)(
19,32)(20,29)(21)(22,25)(23,27)(24)(26)(28 35)(30)(36)(39)(42,43)(47)(48,51)( 50)(52)
for A(4) is found and the new orbits are
Orb(A(4))={1,5,6,40,45,49}{2,9,52}{3,7,41,46,48,51}{4}{8,39,44} {10,42,43}{11,38,47}
{12,21,37}{13,18,34}{14,19,23,27,31,32}{15,20,22,25,29,33}{16,17,36}{24}{26,28,35}
{30}{50} and the order of the stabilizer A(4) is |A(4)|=|A(4,11)|.|Orb(11,A(4))|=2.3=6
($XF_{LMIN}$=13). Then, a check for CS2/CS4 follows (LMIN=2<LP=4). We check for NRO
for each vertex $Z_L \in$ FRPO($X_L$), L=LMIN+1,...,LP=3,4, i.e., if there are changes of the
orders |A(4,13,3)| and |A(4,13,47,31)|. The order of |A(4,13,3)| before $\alpha_7$ is $|A^{-\alpha_7}(4,13,3)|$ =
$|A^{-\alpha_7}(4)|/|Orb(13,A^{-\alpha_7}(4))|/|Orb(3,A^{-\alpha_7}(4,13))|$=2/2/1=1 and after $\alpha_7$ it is
|A(4,13,3)|=|A(4)|/|COrb(13,A(4))|/|COrb(3,A(4))|=6/|13,18,34|/|3,46,51,48|=|6/3/4=0.5.
This difference ($|A^{-\alpha_7}(4,13,3)|$ =1 $\neq$ |A(4,13,3)| = 0.5) shows that the orbit
Orb(3,A(4,13))={3,46,51,48} under A(4,13) is incorrect, it is united orbit, i.e., the case is
CS4. (As we'll see later, the correct orbits are {3,46}{51,48}). Hence, the check for
|A(4,13,47,31)| is not necessary. As the case is CS4 we set LP=4 (not changed), LMIN=LP,
gen(A($X_{LMIN-1}$)=gen(A(4,13,47))={$\alpha_7$}; Orb(A(4,13,47))=cycles of $\alpha_7$ and| A(4,13,47))|=2
(the least multiple of the cycle lengths of $\alpha_7$). We also set $XF_{LMIN}$=31 and we start the
selection of a new vertex from the current $X_{LP}$=14 and since it is the last vertex in SC($\Pi_4$)
we make a move back to the level L=3 selecting the vertex 46. We omit the following
actions of the algorithm. We only mention the last generators
$\alpha_8$=(1,5)(2)(3,51)(4)(6,40)(7,46)(8)(9,52)(10,42)(11,47)(12)(13,18)(14,32)(15,22)(16,36)(1
7)(19,23)(20,33)(21,37)(24)(25,29)(26,35)(27,31)(28)(30)(34)(38)(39,44)(41,48)(43)(45,4
9)(50) (14,27,19) (15,20,25)(16,17,36)(22,29,33)(23,31,32)(24)(26,35,28). The orbits due
to the generators $\alpha_8$ and $\alpha_9$ are:
Orb(A(4))={1,5,6,40,45,49}{2,9,52}{3,7,41,46,48,51}{4}{8,39,44}{10,42,43}{11,38,47}
{12,21,37}{13,18,34}{14,19,23,27,31,32}{15,20,22,25,29,33}{16,17,36}{24}{26,28,35}
{30}{50} and |A(4)|=|Orb($XF_{LMIN}$,A(4))|.|A(4,$XF_{LMIN}$)|=6.1=6, where $XF_{LMIN}$=14. The last
generator found by PART2 is
$\alpha_{10}$=(1,27)(2,28)(3,29)(4,30)(5,31)(6,32)(7,33)(8,34)(9,35)(10,36)(11,37)(12,38)(13,39)
(14,40)(15,41)(16,42)(17,43)(18,44)(19,45)(20,46)(21,47)(22,48)(23,49)(24,50)(25,51)
(26,52). The orbits of A due to the generators $\alpha_8$, $\alpha_9$ and $\alpha_{10}$ are given at the beginning of
this section and |A|=|A(4)||Orb(4,A)|=6*2=12. Thus, the output is: |A|=12, Orb(A) and
generators $\alpha_8$, $\alpha_9$ and $\alpha_{10}$.

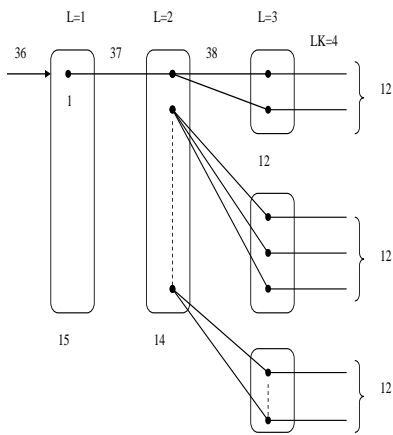

**Figure 17. The search tree for A50 rigid graph**

*5.3.3. Example of a search tree for a rigid graph*

The example is for graph *A50, Mathon* [29]*, n=50,* regular bipartite graph, *k=105=15\*7+35\*3,* **Error! Reference source not found..**)*. The bouquet |B(1)|=168, |SC($\Pi_2$)|=14, each selected cell SC($\Pi_3$) has size |SC($\Pi_2$)|=12. The search tree is full.

**5.4. Correctness and analysis of the algorithm**

We do not prove formally the correctness of the algorithm but it follows from the description of the algorithm. If the bouquets of each representative of an orbit in SC($\Pi_{LP}$) are correct and the whole orbit of each selected vertex in SC($\Pi_{LP}$) is traversed then according to Theorem 4 GOO(A($X_{LP-1}$)) and B(A($X_{LP-1}$)) will be determined correctly. The main problem is to guarantee the correctness of the bouquets - this is shown in the description of the cases CS1, CS2, CS3 and CS4. The traversal of the whole orbit of each selected vertex in SC($\Pi_{LP}$) is also guaranteed (Figure 9). The formal correctness prove of the algorithms and their analysis will be presented in a new article..

*6. Vsep-is algorithm for graph isomorphism(GI algorithm)*

Vsep-is algorithm finds an isomorphism between two graphs G1(n1,k1), G2(n2,k2), if any. It is based on Corollary 3 of theorem 3 and theorem 4. Firstly, it checks for some necessary conditions for isomorphism: a) n1=n2?; b) k1=k2? c) equality of sorted degree sequences of vertices of G1,G2;c) compatibility **of** $\pi1$=RP(G1,$\pi_u$); $\pi2$=RP(G2,$\pi_u$); d) equality of indices of the target cells of $\pi1$ and $\pi2$. If one of these conditions does not hold the graphs are not isomorphic. Otherwise, the bouquet B(x1, G1) is determined by applying PART1 to the first vertex x1∈SC($\pi1$).

Then, by applying PART2 for each vertex y∈SC($\pi2$,G2) is generated one numeration that is compared with the numerations of the **bouquet B(x1,G1) for an** isomorphism, If there is an isomorphism for the current numeration then a message 'G1≈G2'**,** printing the isomorphism and stop follows**.** If all numerations (instruction 6) derived from each vertex

y∈SC(π2,G2) do not form an isomorphism with the bouquet B(x1,G1) then a message that the graphs are not isomorphic follows.

| |
|---|
| **Input:** Graphs G1(n1,k1), G2(n2,k2), |
| **Output:** G1 and G2 are isomorphic or no; print the isomorphism if any |
| 1: **if (n1≠n2 or k1≠k2 or ordered sequences of vertex degrees of G1,G2 are not equal) then** output 'G1!≈G2';**return end if** |
| 2:   Define π1=RP(G1,$\pi_u$); π2=RP(G2,$\pi_u$); **if** π1,π2 are not compatible **then message** 'G1!≈G2'; **return end if** |
| 3. Define SC(π1), SC(π2); **if**(index(SC(π1)) ≠ index(SC(π2)))**then** output 'G1!≈G2'; **return end if** |
| 4. Apply PART1 to the first vertex x1∈SC(π1) to determine the bouquet B(x1,G1) |
| 5. **for** each vertex y∈SC(π2,G2) **do**     determine one numeration *num* by applying SFM(y);     compare *num* with the bouquet B(x1,G1)  for isomorphism;     **if** there is an isomorphism **then** output  'G1≈G2' and the isomorphism; **return end if**   **end for** |
| 6. output 'G1!≈G2' |

**Figure 18. Vsep-is algorithm**

## 7. The heuristic algorithms (Vsep-orb, Vsep-hway, Vsep-sch)

They are based on Theorem 3. For determining whether two vertices *x* and *y* are similar two partial bouquets are built for both vertices and then, some automorphisms between the numerations of these bouquets are determined. To determine certainly that x and y are similar one of the bouquets should be full. Consequently, the probability to find at least one automorphism mapping *x* to *y* is less than 1 if we use the algorithm with partial bouquets. This makes the algorithms inexact. Instead, less time is needed for bouquets building and less storage is needed for them because of their smaller sizes. We describe two heuristic algorithms Vsep-orb, and Vsep-hway in this section.

### 7.1.  TABLE1 and TREE4 procedures

These procedures are needed for the following description of the heuristic algorithms.
TABLE1 heuristic procedure generates in SC(Π) numerations by the operations *fork* and *RRST*. Each new numeration is compared with previously found numerations (stored in HT) for an automorphism. If there is an automorphisn it is added to previously found set of generators. If there is no an automorphism  the numeration is put into HT. RRST tree is defined by the parameters W (width) and D (depth): for each level L, 1< L ≤ 1+D, in SC($\Pi_L$) the number of selected vertices is W. The experiments show that when the selected vertices are evenly distributed in SC($\Pi_L$) of RRST the results are more correct compared with other ways of distribution. One call to TABLE1 procedure with fixed parameters W and D usually is not sufficient to determine a correct GO(A). That's why several calls to TABLE1 with different parameters (W, D) are needed – this is done by TREE4 procedure (Figure 20). The parameters of TREE4 have the following meaning: mbrsyvp – maximal number of successive calls to TABLE1 with no changed norb, wth1 – the start value of W, wth1+br1a - the last value of W, dpth1 – the start value of D, dpth1+br2a - the last value of D. The parameters have default values but there is a possibility the user to change them.

| |
|---|
| *Input*: G- a graph, Π – a stable partition, W, D – a width and a depth of the reduced regular search tree RRST;<br>*Output*: Generators and orbits of A=A(G, Π) stabilizer, norb - a number of orbits;<br>*Local variables*: HT- a hash table of partial bouquets of generated numerations |
| T1: **Initialization:** Set Orb(A(G, Π)) to discrete orbits; gen( A):=∅; HT empty; |
| T2: Find *SC(Π)* target cell; |
| T3: Select each vertex $x \in SC(\Pi)$ that is not similar to a previous vertex in SC(LB) under current orbits. Generate a numeration *n1* for x vertex by a series of forward move and compare it for an automorphism α with the numerations in HT. If there is α then: gen(A) :=gen(A) ∪ {α} and determine new Orb(A(G, Π)) and *norb*. If there is no α then add *n1* into HT. We call this operation ***a fork***. |
| T4: *A **reduced regular search tree (RRST)*** is built for each representative of the orbits in *SC(LB)* determined so far. The orbits are sorted in increasing order of their lengths starting from the smallest one. Each representative of an orbit is a root of a tree. The depth of RRST is D. Each node of the RRST corresponds to a selection level L, $1 < L \leq 1+D$ and a target cell SC(L). In each $SC(\Pi_L)$ a fixed number *W* of vertices *y* are selected, then a forward move to next level L is made for each *y*. In the 1+D level only a series of forward moves SFM is made for each selected vertex until a discrete partition (numeration *n1*) is obtained. The number *W* may be regarded as out-degree of the ***RRST*** node. The vertices in SC ($\Pi_L$) are selected successively by step st= $\|SC(\Pi_L)\| / W \geq 1$ starting from the beginning of *SC* $\Pi_L$), i.e. they are evenly distributed in $SC(\Pi_L)$. To determine an automorphism α and new orbits each numeration *n1* is compared with the numerations in HT table. If there is α then: gen(A) :=gen(A) ∪ {α}, new Orb(A(G, Π)) and *norb* are .determined. If there is no α then add *n1* into HT. The procedure stops when RRSTs are generated for all representatives of the orbits in $SC(\Pi_L)$. |

Figure 19. TABLE1 procedure

| |
|---|
| **Input:** graph G, equitable partition Π, parameters mbrsyvp,wth1, b1a, dpth1, br2a<br>**Output:** GO(A(Π)), norb – number of orbits of A(Π) |
| **1: DO** wth=wth1, wth1+br1a |
| 2:   **DO** dpth=dpth1, dpth1+br2a |
| 3:     TABLE1(Π, WTH, DPTH, norb) |
| 4:     **IF** (norb=1 **or** SC($\Pi_1$) contains one orbit **or** norb remains unchanged after *mbrsyvp* number of successive calls to TABLE1) **then** |
| 5:       exit |
| 6:     end if |
| 7:   END DO |
| 8:END DO |

Figure 20. TREE4 procedure

## 7.2. Vsep-orb algorithm

It has two main steps S1 and S2:
S1) Determine the orbits $Orb(G, \Pi)$ by an heuristic TREE4 algorithm (Figure 20);
S2) Use PART1, PART2 algorithms to determine $GOO(G, \Pi_{ORB})$, where $\Pi_{ORB}$ is a partition which cells to the orbits found in S1 (orbit partition).
TREE4 does not guarantee the computation of the exact orbits of $GOO(G, \Pi)$ and consequently Vsep-orb is an heuristic algorithm - it does not guarantee the computation of the exact $GOO(G, \Pi)$. But if the parameters of Vsep-orb are selected in proper way the probability of the exact computation of $GOO(G, \Pi)$ is very close to 1. The experiments confirm this.

## 7.3. Vsep-sch algorithm

Vsep-sch algorithm has two main steps:
**S1**) Determine the base points, generators and orbits of $Aut(G, \Pi)$ by TABLE1 heuristic algorithm (fig. 18) applied with a parameter LB = 1;
**S2**) Determine the order $|Aut(G, \Pi)|$ using Schreier-Sims algorithm [5,31,32] on the base points and generators found in S1. Obviously, at this step the strong generators (see the definition below) of $Aut(G,\Pi)$ are determined.
Recall [32] that a base for a group A is a sequence of points $B=[\beta_1,\beta_2,\ldots, \beta_m]$ such that the stabilizer $A_{\beta_1,\beta_2,\ldots,\beta_m} = I$. Schreier-Sims algorithm determines a chain of stabilizers
$A=A^{(1)} \geq A^{(2)} \geq \ldots \geq A^{(m)} \geq A^{(m+1)} = I$, where $A^{(j)} = A_{\beta_1,\beta_2,\ldots,\beta_{i-1}}$.
Each stabilizer $A^{(j)}$ is represented by a series of all right cosets to $A^{(j+1)}$
$A^{(j)} = A^{(j+1)} \setminus A^{(j)} = A^{(j+1)} \cup A^{(j+1)} \alpha_2 \cup A^{(j+1)} \alpha_3 \cup \ldots \cup A^{(j+1)} \alpha_n$
The collection of elements in $A^{(j+1)} \setminus A^{(j)}$, j=1,2, . . . ,m are called ***strong generators*** for A toward B. The algorithm [33] is constructing a table T(n,m) of strong generators, where T(i,j) is an element of the group A which fixes $\beta_1, \beta_2,\ldots,\beta_{j-1}$ and maps $\beta_j$ to i. If no such element exists, then the entry is empty. The non-empty elements of the j-th column of T form a set of coset representatives $U^{(j)}$ of $A^{(j+1)}$ in $A^{(j)}$.
The step S2 algorithm (fig. 21) uses a procedure called ***sift*** (fig. 22) that for a given automorpism α compares β(j) with y=α(β(j)), j=1,2,…,m: (a) if y= β(j) then a pass to the next base point β(j+1) follows; (b ) if y ≠ β(j) then a check if α belongs to the stabilizer $A^{(j)}$ follows, i.e. if T(y,j) is empty or no. If T(y,j) is empty then T(y,j) = α, the number of nonempty elements in column j of T is increased by 1, c(j)=c(j)+1, and the ***sift*** ends. If T(y,j)=γ then a new automorphism $α = γα^{-1}$ is determined and all above actions are applied to α for the next base points β(j+1). The procedure *sift* modifies the table T by inserting at most one new coset representative.

Step S2 algorithm is with base points - that is the difference from FHL version of Schreier-Sims method in [33]. When the first 3 instructions of S2 algorithm are finished, the table T should have the property that α ε A iff α can be expressed as $a_1 a_2 \ldots a_m$, where $a_i$ is a member of the j-th column of T. This is called *canonical representation* of α toward B. The time complexity of step S2 is $O(n^6)$ [33,34]

| |
|---|
| **Input:** n, generators, B - base points, m==\|B\| |
| **Local variables:** T(n, m)-2 dimensional integer array, c – integer array of size m, c(j) is the number of nonempty elements in column j of T; Output: Aut(G, Π) |
| **1;** Set each element of *c* to 1;Set each element T(i,j) empty; |
| **2: sift(n,** α,B,m,T,c,**)** {Pass each generator to the procedure sift}; |
| **3: sift(n,**γ,B,m,T,c,**)** {Pass the product γ of each pair of representatives in T to the procedure sift until no new element is inserted in T}; |
| **4:** Determine the order \|Aut(G, Π)\|=$\prod_{j=1}^{m} c(j)$. |

**Figure 21. S2 step of Vsep-sch algorithm**

| |
|---|
| **Input: n,α, B, m, T, c; c-array of size m=\|B\|,c(j) is the number of nonempty elements in column j of T;Output: T,c** |
| **1: for** j:=1 **to** m **do** |
| **2:** y:= α(B(j)) |
| **3: if**(y=B(j)) **cycle** |
| **4: if** (T(y,j) is empty) **then** |
| **5:** T(y,j) := α; c(j) := c(j)+1; **return** |
| **6: else** |
| **7:** γ := **T(y,j);** α := γα$^{-1}$ |
| **8: end if** |
| **9: end for** |

**Figure 22. sift procedure**

## 8. Vsep-a automatic version of Vsep algorithm

Experiments show that Vsep-sch algorithm runs long for graphs with large \|Aut(G)\| (small number of orbits), especially for transitive graphs. In these cases T table is large and full or almost full and evidently this causes more computation. On the other hand, Vsep-orb runs fast on such graphs. This is the reason for developing Vsep-a algorithm that chooses Vsep-orb when the number of orbits is small (≤ 3) and Vsep-sch otherwise. There are also other criteria for this selection. Vsep-a algorithm also chooses Vsep-e algorithm for many graphs on the base of some criteria, for example for some non-regular graphs.

## 9. Program implementation of Vsep algorithms

All proposed algorithms in this paper are implemented in Fortran programs that can be compiled by any Compaq Fortran compiler. The program version published on author's web site section My Programs [35] is called VSEP_PUB4. It implements Vsep-e, Vsep-orb, Vsep-sch and Vsep-a version of VSEP algorithms. The user may choose any version and set different parameters (see user guide).

## 10. Experimental results

In this section we present: a) experiments on almost all benchmark graphs from [19] that compare the performance of VSEP_PUB4 program with Traces (nauty2.6r5 [19]) (Table 4) –one of the most competitive known tool for the worst cases; b) The running times of VSEP_PUB4 for whole families of some benchmark graphs are given in Table 5.
 Some of the experimental results from VSEP _PUB4 (as they are received from the program) are in RESVSE4P_PUB4OBSHT file given in [35]. The most difficult graphs for our algorithms are the graphs with $|Aut(G)|=1$ or with small $|Aut(G)|$. It is known that none of the known algorithms outperform others for all graphs. For each algorithm there are specific difficult and easy graph families. A graph family may be easy for one algorithm and very difficult for another. The same is for Vsep and Traces. Even more, for Vsep different cell selectors give different running times – none of the cell selectors outperform others. The chooser of cell selectors does not always choose the optimal cell selector. For each result we show the cell selector for which it is obtained. The experiments were carried out on a laptop Dell, CPU: Intel(R) Core (TM) i5-3317U@ 1.7 GHz, Memory: 8 GB, OS: 64 bit Microsoft Windows 7 Profesional. For the experiments, we have used all the benchmark graphs of nauty&Traces page [19] that includes a variety of graph families with different characteristics. We show mostly the results for the graphs that are worst cases for either of the compared tools and cell selectors for which they are obtained (only for Vsep). It is evident (See Table 4) that Vsep outperforms Traces considerably for the graphs tnn(39)_1014-1, chh_cc(7-7)_1078-1, f-lex-srg-10-1 and f-lex-srg-50-1 . On the other side Traces outperforms Vsep-e considerably for the graphs of projective planes (pp-16-14, 15, 22, pp-25-90, 116) , had-112, had-176 and latin-sw-112 but for the same graphs Vsep-a outperforms Traces considerably. There are no essential differences between Vsep and Traces on the other graphs on the Table 4! Traces is slow for the graphs with large order of the automorphism group and large number of generators – my experience show that this is may be due to the use of Schreier-Sims method. The main disadvantage of Vsep-e is the storing of the whole bouquet of the first selected vertex-millions of words for some graphs. For almost all benchmark graph families the heuristic versions are many times faster than the exact one and at correctly chosen parameters give correct results.

| G graph, n*, minval*, maxval*, \|Aut(G)\|, norb* | T in seconds | | |
|---|---|---|---|
| | Traces | vsep-e* | vsep-a* |
| | | izb*, bouquet size*.par* | selected version(orb for vsep-orb,shc for vsep-sch,e-for vsep-e),izb*,par* |
| tnn(39)_1014-1, 1014, 4,312, 6.314790834154e174,12 | 5654.38 | 1.01, 1,1,nop<br>0.83, 4,1,nop | 1.03,e,1,nop<br>0.80,e,4,nop |
| cmz-50, 1200,2,5, 1.267650600228e32,8 | 0.08 | 0.14, 1, 1,nop | 0.13,e,1,nop |
| ag-49,4851,49,50, 2.710632960000e10,2 | 0.11 | 0.14, 1,21,nop | 0.14,e,1,nop |
| cfi-200,2000, 3,3, 2.535301200456e30, 800 | 0.14 | 1.17, 1,4,nop<br>1.06,4,4,nop | 1.17,e,1,nop<br>1.07,e,1,nop |
| chh_cc(7-7)_1078-1 1078,3,45,4.907372642035e71,8 | NA | 0.17, 1,1,nop | 0.17,e,1,nop |
| had-128,512,129,129, 1.073807313661e19,1 | 0.02 | 0.11,3,def | 0.11,e,1,def |
| had-112,448,113,113, 1677312,1 | 0.05 | 85.75,1,32287,def | 0.42.e,1,def |
| had-176,704,177,177. 15257088,1 | 0.14 | 429.75,1,21210,def | 1.88,e,1,def |
| had-256,1024,257,257, 1.401962828716e24,1 | 0.05 | 0.61,1,1,def | 0.62,e,1,def |
| latin-30.dre,900 ,87,87, 43200.1 | 0.03 | 0.71,1,1106,def<br>0.66,4,962,def | 1.33.e,1,def |
| latin-sw-30-1,900 ,87,87, 1,900 | 0.17 | 4.33, 1,812; 1,2,1,1,1 | 3.16,e,1; 1,2,1,1,1 |
| had-sw-112,448,113,113, 2,224 | 1.61 | 179.7,1, 519624; 1,2,1,1,1 | 16.69,sch,1,def<br>0.61,sch, 1,2,1,1,1 |
| lattice-30.dre, 900 , 58,58, 1.407181592771e65,1 | 0.05 | 0.09, 1 ,1,def | 0.12,orb,1,def |
| 10cube,1024,10,10, 3.715891200e9,1 | 0.02 | 0.11, 1,1,def | 0.11,orb, 1,def |
| paley-461.dre ,461,230,230, 1.06030e5,1 | 0.02 | 0.14, 1,1,def | 0.15,orb, 1,def |
| pg2-49, 4902,50,50, 1.328752276992e14,1 | 0.11 | 2.00, 1,21,def | 2.00,orb,1,def |
| pp-16-14,546,17,17, 2304,14 | 0.66 | 55.24, 1, 420960; | 0.44,sch,1; |

| G graph, n*, minval*, maxval*, \|Aut(G)\|, norb* | T in seconds | | |
|---|---|---|---|
| | Traces | vsep-e* | vsep-a* |
| | | izb*, bouquet size*.par* | selected version(orb for vsep-orb,shc for vsep-sch,e-for vsep-e),izb*,par* |
| | | 3,3,2,1,1<br> 8.50, 2, 52680;<br>3,3,2,1,1 | 3,3,2,1,1 |
| pp-16-22,546,17,17, 9216,10 | 0.55 | 15.4, 1,117264;<br>3,3,2,1,1<br>2.26,2,13336;<br>3,3,2,1,1 | 0.26,sch,1;<br>3,3,2,1,1 |
| pp-25-1,1302,26,26, 609336000000,2 | 0.05 | 0.12,2,10;<br>3,3,2,1,1 | 0.12,orb,1,<br>3,3,2,1,1 |
| pp-25-90,1302,26,26,1000,40 | 42.62 | 242.6, 2, 522995,<br>3,3,2,1,1;<br>2244.03,1, 5016276,def | 9.7,sch,1;<br>3,3,2,1,1 |
| pp-25-116,1302,26,26, 500,64 | 104.82 | 500.0, 2 , 1058568,<br>3,3,2,1,1 | 6.00,sch,1,<br>3,3,2,1,1 |
| pp-27-10, 1514,28, 28, 122472,6 | 1.05 | 40.0,,2, 73008<br>3 5 2 1 1 | 4.59,sch,2;<br>3 5 2 1 1 |
| pp-49-1,4902,50,50,288120,10 | 236.87 | 5631.0, 2, 1975680;<br>2,5,2,2,1<br>5600.0, 4, 1975680<br>2,5,2,2,1<br>5529.05, 2, 1975680<br>2,5,2,2,1 | 117.4,1;<br>2,5,2,2,1;<br>90.1,2;<br>2,4,2,2,1 |
| mz-aug-50,1000,3,6, 10141204801825835211973625643008,250 | 0.00 | 0.22,1,1,def | 0.22, orb,1,def |
| mz-aug2-50,1200,2,5, 50706024009129176059868128215 04,700 | 0.00 | 0.54,1,1,def | 0.53,orb, 1,def |
| mz-50,1000,3,3, | 0.00 | 0.22,1,1,def | 0.20,orb, 1,def |

| G graph, n*, minval*, maxval*, \|Aut(G)\|, norb* | T in seconds | | |
|---|---|---|---|
| | Traces | vsep-e* | vsep-a* |
| | | izb*, bouquet size*.par* | selected version(orb for vsep-orb,shc for vsep-sch,e-for vsep-e),izb*,par* |
| 1014120480182583521197362564300 8,250 | | | |
| f-lex-reg-50-5,1700, 8,30,16384,1016 | 0.06 | 0.42,1,2,def | 0.42,orb,1,def |
| f-lex-srg-10-1,790,16,75,16,,439 | NA | NA | 0.17,sch,2; 2 2 1 2 2 |
| f-lex-srg-50-5,3950, 16,75,32768,2360 | NA | NA | 10.69,sch,2; 2 2 1 2 2 |

**Table 4.** Experimental comparison of VSEP4P_PUB4 program with Traces. n - number of vertices, norb-number of orbits, vsep-e - exact version, vsep-a – automatic version, izb - the numbering of the chosen cell selector; minval, maxval - minimal and maximal degree of a vertex of the graph, bouquet size – number of the stored numerations; PAR – parameters mbrsyvp, wth1, dpth1, br1a, br2a; 4,5,2 ,1,1 are default parameters (def=default),nop-no parameters since there is no call to tree4;NA-not attended

| Graph family | Parameters: *mbrsyvp*, *wth1, dpth1, br1a, br2a; 4,5,2 ,1,1-default (def)* | izb | T[secs]-running time |
|---|---|---|---|
| mathon | Def | 1 | 0.1248 |
| mathon dobling | Def | 1 | 0.312 |
| tnn | Def | 1 | 3.97 |
| tnn | Def | 4 | 3.31 |
| cmz | Def | 1 | 2.37 |
| ag | Def | 1 | 0.71 |
| chh | Def | 1 | 2.19 |
| cfi | Def | 1 | 38.94 |
| cfi | Def | 4 | 36.77 |
| latin | Def | 1 | 3.79 |
| lattice | Def | 1 | 0.99 |
| hypercubes3 | Def | 1 | 0.25 |
| paley | Def | 1 | 1.90 |
| ppsmall | Def | 1 | 0.56 |
| pp16 | 3 3 2 4 1 | 1 | 7.28 |
| pp16 | 3 3 2 2 1 | 1 | 7.22 |
| pp16 | 3 3 2 1 1 | 1 | 7.19 |
| pp25 | 3 3 2 1 1 | 1 | 746.11 |
| pp27 | Def | 2 | 60.66 |
| pp27 | 3 5 2 1 1 | 2 | 57.60 |

| Graph family | Parameters: *mbrsyvp*, *wth1, dpth1, br1a, br2a; 4,5,2 ,1,1-default (def)* | izb | T[secs]-running time |
|---|---|---|---|
| pp27 | 3 5 2 1 1 | 3 | 58.26 |
| pp27 | 3 4 3 4 2 | 1 | 68.53 |
| pp27 | 3 4 3 4 2 | 2 | 88.54 |
| pp27 | 2 4 3 4 2 | 2 | 70.76 |
| pp27 | 3 4 3 3 2 | 2 | 88.90 |
| pp27 | 3 4 3 4 1 | 2 | 60.31 |
| pp27 | 2 4 3 3 2 | 2 | 76.32 |
| pp27 | 2 4 3 3 1 | 2 | 68.62 |
| pp27 | 2 4 3 4 1 | 2 | 75.48 |
| pp27 | 2 4 3 2 1 | 2 | 69.28 |
| pp27 | 2 4 3 1 1 | 2 | 68.82 |
| pp49 | 3 5 2 2 1 | 1 | 505.00 |
| pp49 | 2 5 2 2 1 | 2 | 349.60 |
| pp49 | 2 5 2 2 1 | 1 | 365.10 |
| had | 4 5 2 1 1 | 1 | 293.28 |
| had | 4 5 2 1 1 | 3 | 278.40 |
| had | 3 6 2 1 1 | 3 | 281.36 |
| f-lex=srg | 2 2 1 2 2 | 2 | 589.29 |
| f-lex=reg | Def | 1 | 18.47 |
| pg | Def | 1 | 7.84 |
| pg | 3 4 2 1 1 | 1 | 7.80 |
| pg | 3 3 2 1 1 | 1 | 7.80 |
| mz-aug | Def | 1 | 2.09 |
| mz-aug2 | Def | 1 | 3.77 |
| total run time for all cases in the table | | | 4478.15 |

**Table 5.** Experimental results from VSEP4P_PUB4 program for whole families of the most of the benchmark graphs in [19], some of the results are in RESVSEP4OBSHT file [35], izb is the numbering of the chosen cell selector

## 11. Concluding remarks and open problems

Five new algorithms, named Vsep, are described. Four of them are for determining the generators, orbits and order of an undirected graph automorphism group:. Vsep-e – exact, Vsep-orb and Vsep-sch – heuristic and Vsep-a automatically selects the optimal version among Vsep-e, Vsep-orb and Vsep-sch. The fifth algorithm, Vsep-is, is for finding an isomorphism between two graphs. A new approach is used in the exact algorithm: if during its execution some of the searched or intermediate variables obtain a wrong value then the algorithm continues from a new start point losing some of the results determined so far (cases CS3, CS4). The new start point is such that the correct results are obtained.
A new code, named S-code, of a partition of the graph vertices is proposed. S-code is used for reducing the time of comparing the partitions in the proposed algorithms.
The experiments show that the worst case time complexity of Vsep-e for an arbitrary graph is exponential but for some classes it is polynomial. The main difference of the exact one and the widely known tools is the storing of the whole bouquet of numerations of the first selected vertex in the first level. Five cell selectors are used in the algorithms and a chooser of optimal cell selector is presented. Some of the cell selectors are new, namely *mxvectchval* and *mxprchval*.

Experimental comparison of the proposed algorithms with Traces algorithm is made - it shows their worst and best cases. A disadvantage of Vsep-e algorithm is its higher requirements for memory (for some worst cases several millions of numbers are stored. The worst cases for the algorithm Vsep-e are the graphs with smaller order |Aut(G)|, especially the rigid graphs. Vsep-orb and Vsep-sch heuristic algorithms are extremely fast (with some exceptions) compared with the exact one and are almost exact - for all tested thousands of graphs they give correct results. Practically, their requirements for memory are very small. Heuristic algorithms are an important approach for solving so hard problem as graph isomorphism is.

The future work on developing Vsep algorithms will include: a) search for a new cell selector that will reduce the size of the search tree; b) search for a new chooser of a cell selector; c) conduct a comparison of Vsep algorithms with other known GOO algorithms; d) develop parallel versions of Vsep algorithms. Open problems are: i) Develop a new chooser of cell selector that given a graph can determine the best cell selector (that yields a minimal running time of the algorithm); ii) Develop an algorithm that given a graph and the first numeration can determine the parameters *mbrsyvp*, *wth1, dpth1, br1a, br2a* of the forest of reduced regular trees such that when used the heuristic versions of VSEP algorithms will give correct results for a minimal running time; iii) Develop an algorithm that given a graph on the base of some criteria to choose the optimal algorithm from the known published algorithms.

Acknowledgements

The author thanks to Brendan McKay, Bill Kocay, Rudi Mathon, Vladimir Tonchev, Ulrich Dempwolff, Gordon Royle, Adolfo Piperno, Jose Luis López-Presa, Petteri Kaski and Hadi Katebi for the extensive discussions on the graph isomorphism problem and exchanging graphs with interesting properties and papers on this problem. Special thanks to my former diploma student Apostol Garchev who contributed in performing and repeating some of the experiments on the developed algorithms. The author would like to thank all colleagues and students who contributed to this study.